\documentclass{iau}

\pdfoutput=1

\usepackage{amsmath}
\usepackage{graphicx}
\usepackage{multirow}

\usepackage{amsfonts}
\usepackage{rotating}
\usepackage{natbib}
\usepackage{txfonts}
\usepackage{color}
\usepackage{wrapfig}
\usepackage{color}
\usepackage{mhchem}
\usepackage{colortbl}

\begin{document}

\lefttitle{T. Shimonishi}
\righttitle{Chemistry of Low-Metallicity Star-Forming Regions}



\title{The Role of Metallicity in the Chemical Evolution of Star-Forming Regions}

\author{Takashi Shimonishi}
\affiliation{Institute of Science and Technology, Niigata University, Ikarashi-ninocho 8050, Nishi-ku, Niigata 950-2181, Japan}

\begin{abstract}
Understanding the chemistry of the interstellar medium in low-metallicity environments is crucial to unveil physical and chemical processes in the past Galactic environment or those in high-redshift galaxies, where the metallicity was significantly lower compared to the present-day solar neighborhood. 
This is also important for the understanding of the diversity of the chemical evolution in various regions of our Galaxy. 
Nearby low-metallicity laboratories, such as the outskirts of our Galaxy, the Large and Small Magellanic Clouds, and the other gas-rich dwarf galaxies in the Local Group, will provide important insights for this purpose. 
In the last decade, there has been great progress in astrochemical studies of interstellar molecules in low-metallicity star-forming regions. 
Single-dish radio observations have detected various dense gas tracers in these regions, which have revealed the molecular-cloud-scale ($>$several pc) chemistry at low metallicity. 
With ALMA, emission from dense and high-temperature molecular gas associated with protostars (i.e., hot molecular cores) are detected in the LMC, SMC, and outer Galaxy, which have revealed the chemical complexity of star-forming cores ($<$0.1 pc) at low metallicity. 
Besides gas-phase species, infrared observations have revealed chemical compositions of ices around deeply embedded protostars in the LMC and SMC. 
Do molecular abundances simply scale with the metallicity? 
If not, which processes govern the chemistry in the low-metallicity interstellar medium? 
In this proceeding, I will discuss the role of metallicity in the chemical evolution of star-forming regions based on recent observations of interstellar molecules in low-metallicity environments. 
\end{abstract}


\maketitle

\section{Introduction}

\textit{``Half of the stellar mass observed today was formed before a redshift z $=$ 1.3. 
About 25$\%$ formed before the peak of the cosmic star-formation rate density, and another 25$\%$ formed after z $=$ 0.7."}
As mentioned in the beginning of \citet{Mad14} and also in many other studies, the past universe was active in star formation compared to the present-day universe. 
In addition, the past universe was poor in heavy elements \citep[e.g.,][]{Del19}, because they are synthesized in the stellar interior and the cycle of stellar births and deaths contributes to the enrichment of \textit{``metals"} in the interstellar medium (ISM). 
Therefore, the dominant mode of star formation (and possibly planet formation) as well as the relevant physical/chemical processing of the ISM in the cosmic history should have occurred in low-metallicity environments. 
This is the reason why we need to care about the physics and chemistry of the low-metallicity star-/planet-forming regions. 

More than 300 interstellar molecules including complex organic molecules (COMs) are reported, and now we know that many of the star-/planet-forming regions in the present-day solar neighborhood are chemically rich. 
Then, how was it like in the past low-metallicity universe?
Since most interstellar molecules contain heavy elements, their abundances could be simply lowered by the reduced metallicity. 
Indirect effect caused by the reduced abundance of dust grains may also be important. 
The reduced dust abundance will affect the radiation environment of molecular clouds and protostellar envelopes, which may enhance photochemical reactions. 
For many COMs, grain surface reactions are known to be their dominant formation routes \citep[e.g.,][and references therein]{Wat02, Gar06, Her09}, thus the low dust abundance may reduce the chemical complexity of the ISM. 

\begin{figure*}[t!]
 \begin{center}
  \includegraphics[width=12.5cm]{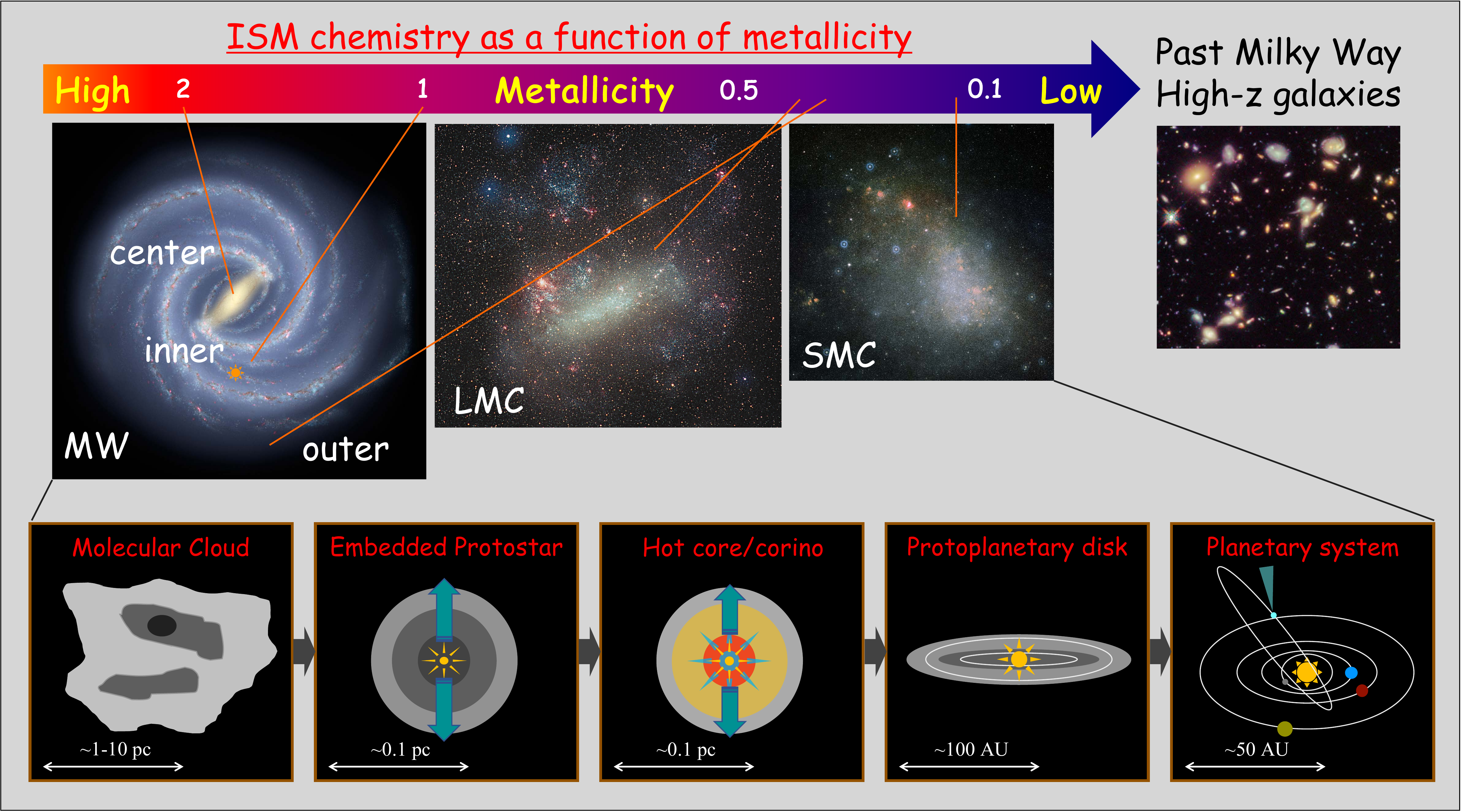}
  \caption{Conceptual diagram of star and planet formation in a nearby low-metallicity laboratory.
  The figure includes a derivative work of the following sources (NASA/JPL-Caltech/ESO R. Hurt; ESA/Hubble E. Slawik; ESA/Hubble and DSS2; NASA/ESA R. Ellis and the HUDF 2012 Team). 
  }
  \label{fig_Concept}
 \end{center}
\end{figure*}

Nearby low-metallicity regions provide us with an excellent laboratory to investigate various astrophysical and astrochemical phenomena that proceed in low-metallicity environments (Fig. \ref{fig_Concept}). 
Our neighborhood star-forming dwarf galaxies, the Large and Small Magellanic Clouds (LMC/SMC), as well as the outer part of our Galaxy\footnote{Here the outer Galaxy is defined as having a galactocentric distance larger than 13.5 kpc. That with a galactocentric distance larger than 18 kpc is sometimes called the extreme outer Galaxy.}, 
are representative of such regions. 
Unfortunately, high-$z$ galaxies are too distant to spatially resolve individual star-forming clouds/cores contained in each galaxy. 
The metallicities of the LMC, SMC, and the outer Galaxy are known to be lower than the solar value by a factor of about two to ten \citep[e.g.,][]{Rus92,Rol02,Hun05,Fer17,Are20}. 
Such a metallicity range roughly corresponds to the mean metallicity of the ISM in the galaxies at $z$ = 0.6--1.8 (6--10 Gyr lookback time), which is close to the peak epoch of star formation in the universe \citep[e.g.,][]{Bal07,Raf12}. 
Besides their unique metallicity environments, thanks to the proximity of these regions \citep[$\sim$10-20 kpc to the outer Galaxy, 50 kpc to the LMC, 62 kpc to the SMC; e.g.,][]{Pie13,Gra14}, we can spatially resolve individual star-forming cores ($\sim$0.1 pc in size), which is essentially important for probing complex molecules in star-forming regions. 

In the last decade, there has been great progress in astrochemical studies of interstellar molecules in nearby low-metallicity systems at various spatial scales of star formation. 
Single-dish radio observations have detected various high-critical-density molecules in these regions, which have revealed the molecular-cloud-scale ($>$several pc) chemistry at low metallicity. 
The Atacama Large Millimeter/submillimeter Array (ALMA) has detected hot molecular cores in these regions, which have revealed the chemical complexity of star-forming cores ($<$0.1 pc) at low metallicity. 
Infrared observations have revealed chemical compositions of ices seen towards deeply embedded protostars in the LMC and SMC. 
In this proceeding, I will discuss the role of metallicity in the chemical evolution of star-forming regions based on recent radio and infrared observations of star-forming objects at various spatial scale (i.e. from molecular clouds to protostellar cores). 

\begin{table}[btp!]
 \centering
 \caption{Summary of single-dish observations of dense gas in low-metallicity molecular clouds}\label{Tab_MolC}
 {\tablefont\begin{tabular}{@{\extracolsep{\fill}} l c p{10em} p{4em} p{6em} c p{15em}}
    \midrule
    Galaxy & Metallicity      & Detected Molecule & Frequency & Telescope & No. & Reference \\
                & (Z$_{\odot}$) &                & (GHz)         &                  &        &  \\
    \midrule
    M33         & $\sim$0.5
                & \ce{HCO+}, \ce{HCN}, \ce{CCH}
                & 85-115 & NRO, IRAM & 8
                & \citet{Ros11, Buc13, Bra17} \\
    LMC         & $\sim$0.4
                & \ce{HCO+(D,$^{13}$C)}, \ce{HCN/HNC}($^{13}$C,$^{15}$N), \ce{CN}, \ce{CCH}, \ce{c-C3H2}, \ce{CS}($^{13}$C,$^{34,33}$S), \ce{SO}, \ce{H2S}, \ce{H2CS}, \ce{SO2}, \ce{H2CO}, \ce{CH3OH}, \ce{CH3CCH}, \ce{N2H+}
                & 85-180, 220-230, 325-365 & SEST, Mopra, APEX, ASTE & $>$20
                & \citet{Joh91,Joh94, Chin96, Chin97, Hei97, Hei99, Wan09, Par14, Par16, Nis16a, Gal20, Gon23} \\
    IC10        & $\sim$0.3
                & \ce{HCO+}, \ce{HCN}, \ce{HNC}, \ce{CCH}, \ce{CS}, \ce{SO}
                & 85-115 & NRO, IRAM & 3
                & \citet{Nis16b, Bra17, Kep18} \\
    NGC6822     & $\sim$0.3
                & \ce{HCO+}, \ce{HCN}, \ce{CCH}, \ce{CS}, \ce{SO}
                & 85-115 & IRAM & 1
                & \citet{Bra17} \\
    SMC         & $\sim$0.2
                & \ce{HCO+}, \ce{HCN}, \ce{HNC}, \ce{CN}, \ce{CCH}, \ce{c-C3H2}, \ce{CS}, \ce{SO}, \ce{H2CO}
                & 85-115, 170-180, 325-365 & SEST, Mopra, APEX, ASTE & 8
                & \citet{Chin97, Chin98, Hei99, Gal20} \\
    \midrule
    Outer MW    & $\sim$0.1--0.5
                & \ce{HCO+}($^{13}$C), \ce{HCO}, \ce{HCN/HNC}($^{13}$C,$^{15}$N), \ce{CN}, \ce{NH3}, \ce{CS}, \ce{SO}, \ce{CCS}, \ce{C4H}, \ce{c-C3H2}, \ce{HCS+}, \ce{H2CO}, \ce{CH3OH}, \ce{CH3CCH}
                & 10-30, \hspace{8pt}85-155 & ARO, IRAM, Effelsberg, TRAO, PMO & $>$40
                & \citet{Ruf07, Bra07, Bla08, Ber21, Fon22a, Fon22b, Col22, Pat22, Bra23} \\
    \midrule
    \end{tabular}}
\tabnote{\textit{Notes}: Only single-dish observations are summarized. 
The detected isotopologues are indicated in parenthesis. 
CO and its isotopologues are detected in all of the sources and not shown in the table. 
Maser observations are not included. 
The sixth column ``No." indicates the total number of clouds with dense gas detection. }
\end{table}

\section{\textbf{Molecular Clouds at Low Metallicity}}
Molecular clouds serve as initial conditions for the subsequent chemical evolution in star formation. 
The chemistry at this stage (typically, $n_{\ce{H2}}$ $\sim$10$^3$--10$^4$ cm$^{-3}$, $T$ $\sim$10--30 K, size $\gtrsim$several pc) is mainly dominated by cold gas-phase chemical reactions. 

Table \ref{Tab_MolC} summarizes single-dish observations of dense molecular gas in various low-metallicity environments. 
Note that, to focus on the molecular complexity, the table does not include observations CO and its isotopologues as well as maser observations. 
Pioneering works on astrochemical studies of low-metallicity molecular clouds are reported in the early 1990's based on observations of the LMC with the 15-m Swedish-ESO Sub-millimetre Telescope (SEST) \citep[e.g.,][]{Joh91}. 
Since then, multiline observations of molecular clouds are extended to many other low-metallicity regions. 
Beside the LMC, SMC, and the outer Galaxy, the target regions include star-forming low-metallicity dwarf galaxies in the Local Group such as IC10, M33, and NGC6822 (see Table. \ref{Tab_MolC} for details). 

\begin{figure*}[tbp!]
 \begin{center}
  \includegraphics[width=12.5cm]{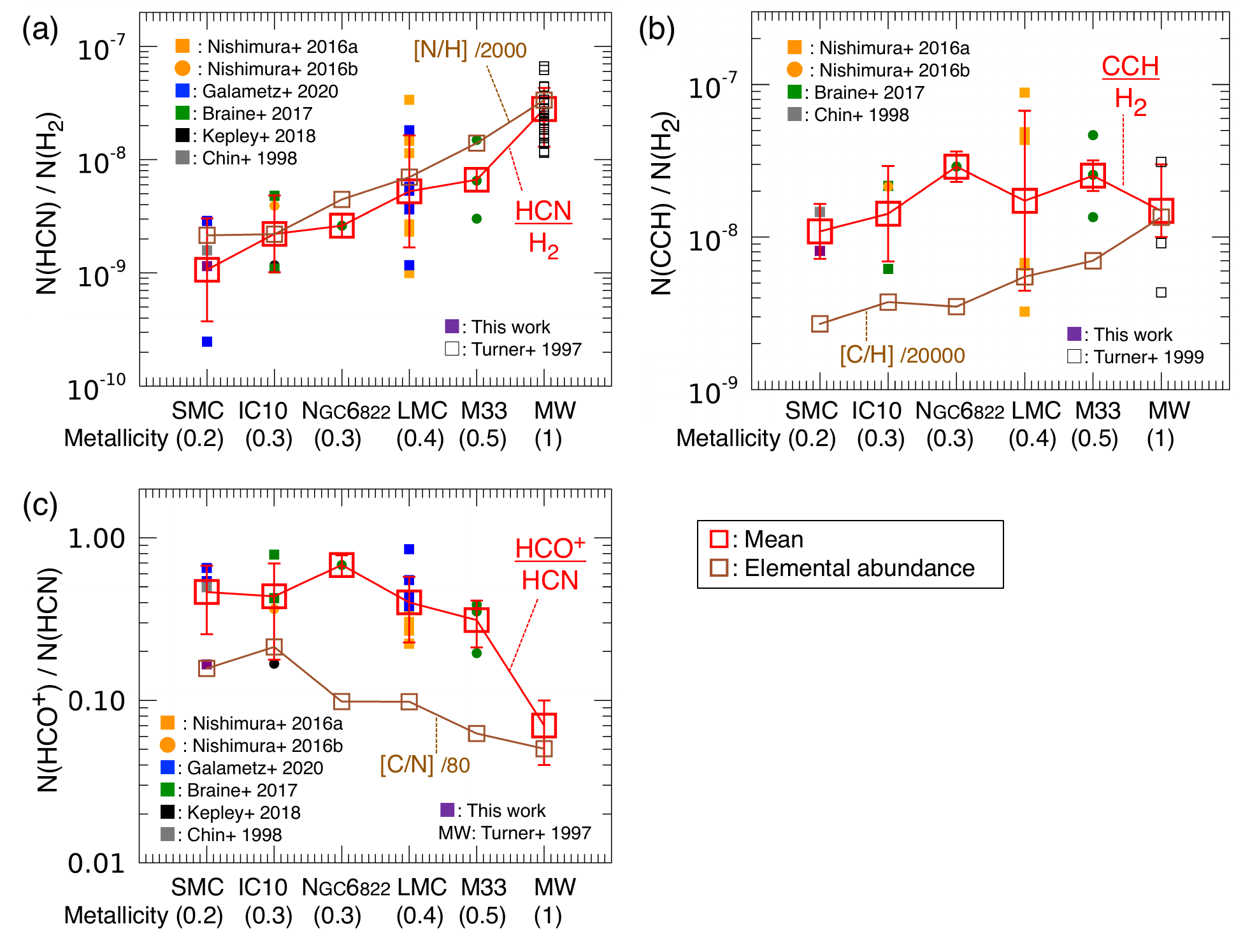}
  \caption{Molecular abundances or abundance ratios of dense clouds in various metallicity environments observed by single-dish telescopes; (a) HCN/\ce{H2}, (b) CCH/\ce{H2}, and (c) \ce{HCO+}/HCN. 
  The red open squares represent the mean value in each galaxy bin, and the error bar represent the standard deviation. 
  Molecular data are adopted from \citet{Tur97, Tur99, Chin98, Nis16a, Nis16b, Bra17, Kep18, Gal20}, while elemental abundances are from \citet{Ski89, Rus92, Mag09a, Mag09b}.   
  The purple data point is calculated in this work based the submillimeter spectral line survey data of a SMC's molecular cloud
  obtained by the ASTE 10 m telescope. 
  }
  \label{fig_MolC1}
 \end{center}
\end{figure*}

Figure \ref{fig_MolC1} shows fractional abundances or abundance ratios of the selected species observed in low-metallicity molecular clouds. 
As a comparison sample of solar-metallicity counterparts, we here use the abundance data of Galactic translucent clouds \citep[e.g.,][]{Tur97}. 
Detailed discussions about the similarity between $>$10-pc-scale spectra of extragalactic molecular clouds and Galactic translucent clouds can be found in \citet{Nis16a,Nis16b}. 

One of the straightforward effect of metallicity on the chemical compositions of molecular gas is ``metallicity scaling effect". 
Such simple scaling effect is actually seen in one of the commonly-used dense gas tracers, HCN. 
Figure \ref{fig_MolC1}(a) plots the fractional abundance of HCN (relative to \ce{H2}) against the metallicity of the host environments. 
The figure also shows the elemental abundances of nitrogen in each region. 
The decrease of HCN abundance in low-metallicity molecular clouds follow the decrease of elemental nitrogen abundances, which can be interpreted as the metallicity scaling effect. 

In contrast, the abundance of a molecular radical, CCH, does not follow such a metallicity scaling. 
Figure \ref{fig_MolC1}(b) shows the abundance ratios of \ce{CCH}/\ce{H2} according to the metallicity. 
Low-metallicity clouds show the higher CCH fraction compared to the solar-metallicity clouds. 
Such a behavior cannot be accounted for by the variation of the elemental carbon abundances as shown in the figure. 
CCH is known to be produced efficiently in photodissociation regions (PDRs) by the photochemistry of parental carbon-bearing molecules \citep[e.g.,][]{Pet05,Mar14}. 
The increase of CCH in low-metallicity molecular clouds may be the consequence of the increased volume fraction of PDRs in low-metallicity environments, which may be caused by the reduced dust abundance. 
Such a picture is consistent with the bright \ce{C+} emission line observed in low-metallicity ISM \citep[e.g.,][]{Mad19}. 

A behavior of molecular abundances that does not follow the elemental abundance pattern is also seen in the \ce{HCO+}/\ce{HCN} ratio, where the \ce{HCO+} is a major molecular ion in clouds (Figure \ref{fig_MolC1} panel c). 
The increase of the \ce{HCO+}/\ce{HCN} ratio in low-metallicity molecular clouds does not follow the elemental C/N ratio as shown in Figure \ref{fig_MolC1}(c). 
An integrated intensity ratio of \ce{HCO+}(1-0) over \ce{HCN}(1-0) is typically $\sim$1 in solar-metallicity molecular clouds, while it is $\sim$1--10 in low-metallicity clouds \citep[e.g.,][]{Nis16a,Nis16b,Bra17,Gal20}. 
Both molecules are often used to trace dense gas, particularly in extragalactic studies, but \ce{HCO+} could be a brighter molecular tracer of dense gas in the case of low-metallicity systems. 

As shown in this section, the effect of metallicity on the molecular-cloud-scale ($>$several pc) chemistry is twofold. 
One is the metallicity scaling effect, where the molecular abundance simply scales with the elemental abundance. 
Another is indirect effect caused by the reduced dust abundance, which can enhance the photochemistry by changing the radiation environment of molecular clouds.

\section{Embedded Protostellar Cores at Low Metallicity}
\begin{figure*}[btp!]
 \begin{center}
  \includegraphics[width=9cm]{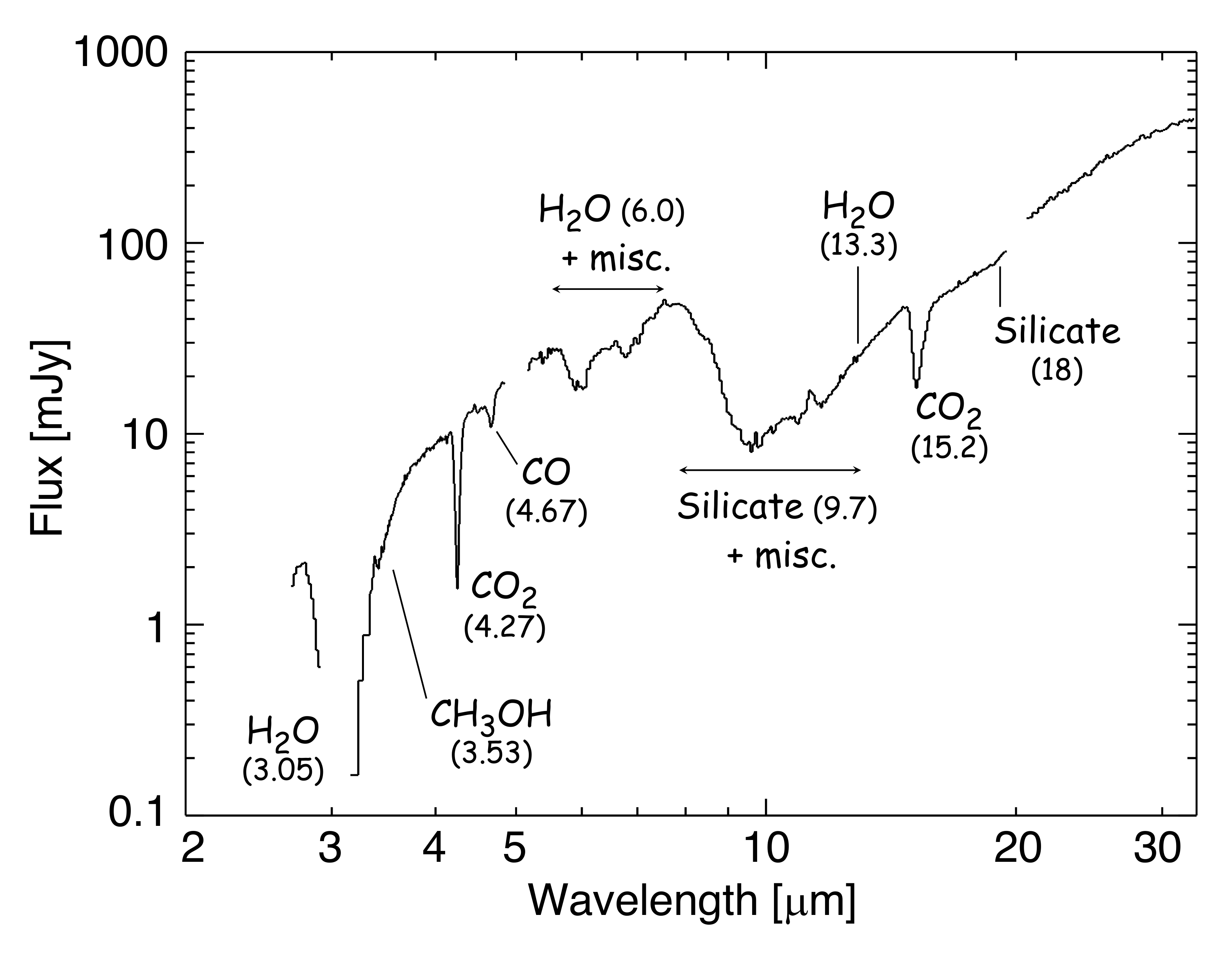}
  \caption{Infrared spectrum of the embedded high-mass protostellar object, ST6, located in the LMC \citep{ST16}. 
  The detected ice/dust absorption bands are labeled with their wavelengths. }
  \label{fig_LMCIce}
 \end{center}
\end{figure*}

Particularly dense parts of molecular clouds are the cocoon of star formation. 
Low-temperature grain-surface reactions largely contribute to the chemical evolution of molecular cloud cores or early-stage protostellar cores (typically, $n_{\ce{H2}}$ $\sim$10$^5$--10$^6$ cm$^{-3}$, $T$ $\sim$10 K, size $\lesssim$0.1 pc), thus infrared observations of ices play a crucial role. 
Figure \ref{fig_LMCIce} shows an example of the ice absorption bands detected towards a high-mass protostellar object in the LMC. 

\begin{table}[btp!]
 \centering
 \caption{Summary of ice detection toward high-mass protostellar objects in the Magellanic Clouds}\label{Tab_Ice}
 {\tablefont\begin{tabular}{@{\extracolsep{\fill}} l c c p{10em} p{24em}}
    \midrule
    Wavelength & Galaxy  & No. & Ice species & Reference \\
    \midrule
    \multirow{2}{*}{2--5 $\mu$m (AKARI, VLT)} 
    & LMC      & 20 &\ce{H2O}, \ce{CO2}, \ce{CO}, \ce{CH3OH}, \ce{XCN}?  & \citet{vanL05, Oli06, Oli11, ST, ST10, ST13, ST16} \\
    & SMC      & 12 &\ce{H2O}, \ce{CO2}                                  & \citet{vanL08, Oli11, Oli13, ST12, ST18} \\
    \midrule
    \multirow{2}{*}{5--20 $\mu$m (Spitzer)} 
    & LMC   & $>$50 &\ce{H2O}, \ce{CO2}, \ce{HCOOH}?, \ce{H2CO}?, \ce{NH4+}?  & \citet{vanL05, Oli09, Oli11, Sea09, Sea11, ST16} \\
    & SMC      & 15 &\ce{H2O}, \ce{CO2}                                       & \citet{Oli11, Oli13} \\
    \midrule
    \multirow{2}{*}{60--70 $\mu$m (Spitzer)} 
    & LMC      & 5 &\ce{H2O} & \citet{vanL10} \\
    & SMC      & 1 &\ce{H2O} & \citet{vanL10_b} \\
    \midrule
    \end{tabular}}
\tabnote{\textit{Notes}: ``?" indicates the tentative detection. The third column ``No." indicates the total number of protostellar objects with ice detection. }
\end{table}

Table \ref{Tab_Ice} summarizes the detection of ice absorption bands towards low-metallicity protostellar cores. 
The first detection of ices associated a low-metallicity protostar is reported by \citet{vanL05} based on observations of an infrared source in the LMC. 
Since then, many ice observations have been reported for embedded protostellar objects (mainly high-mass sources) in the LMC and SMC. 
The near-infrared (2--5 $\mu$m) spectral coverage provided by the InfraRed Camera \citep[IRC,][]{TON07} on board the \textit{AKARI} satellite \citep{Mur07} and the Infrared Spectrometer And Array Camera \citep[ISAAC,][]{Moo98} on the Very Large Telescope (VLT) contribute to the overall abundance analyses of major ice species such as \ce{H2O}, \ce{CO2}, \ce{CO}, and \ce{CH3OH} \citep[e.g.,][]{Oli13,ST16}. 
On the one hand, the mid-infrared (5--20 $\mu$m) spectral coverage provided by the InfraRed Spectrograph \citep[IRS,][]{Hou04} on board the \textit{Spitzer} Space Telescope \citep{Wer04} contribute to the profile analyses of the 15.2 $\mu$m \ce{CO2} ice band, which is sensitive to the thermal processing of ices \citep[e.g.,][]{Sea11}. 
Tentative detection of a far-infrared "emission" band possibly due to the crystalline \ce{H2O} ice is also reported for protostellar objects in the LMC/SMC \citep[e.g.,][]{vanL10}.  

\begin{figure*}[tb!]
 \begin{center}
  \includegraphics[width=10cm]{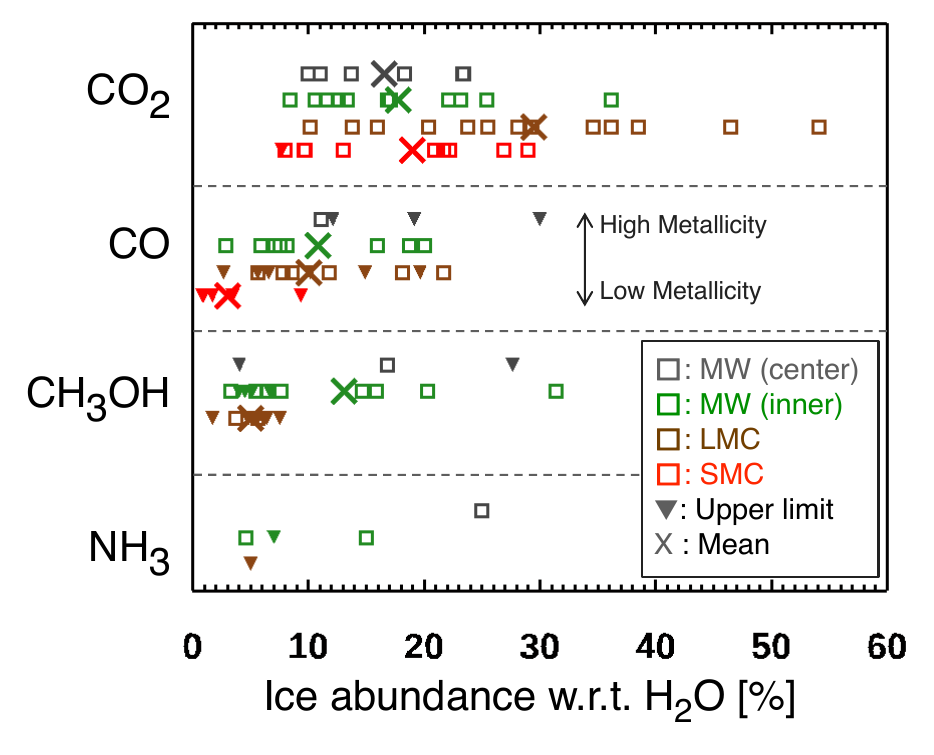}
  \caption{Ice compositions in the envelope of high-mass protostellar objects located in various metallicity environments from the Galactic center to the Magellanic Clouds.   
  The cross symbols indicate the mean value in each regional bin. 
  The plotted data are adopted from MW (center): \citet{Chi00, Chi02, An09, An11}, MW (inner): \citet{Dar02, Gib01, Gib04, Boo15}, LMC/SMC: \citet{Oli13, ST16}. 
  No data is available for the outer Galaxy. 
  }
  \label{fig_IceAbu}
 \end{center}
\end{figure*}

Figure \ref{fig_IceAbu} compares the ice abundances observed in embedded high-mass protostars located in the Galactic Center, the inner Galaxy, the LMC, and the SMC, where the metallicity decreases from the former to the latter. 
The plotted molecular abundances are normalized by the H$_2$O ice column density, since H$_2$O is the most abundant species in ice mantles. 
Naively thinking, the plotted data of ices in different metallicity environments should distribute similarly if the molecular abundances simply scale with the metallicity.

However, the ice abundances in the figure do not follow such a simple metallicity scaling law. 
For example, compared to those of Galactic counterparts, the scatter of the LMC's \ce{CO2} ice abundance is large and the mean \ce{CO2} ice abundance is higher by a factor of about two. 
In contrast, the \ce{CH3OH} ice abundance in the LMC is lower than those of Galactic sources. 

In dense clouds, both CO$_2$ and CH$_3$OH are mainly produced by grain surface reactions \citep[e.g.,][]{Tie82}. 
The surface reaction of $CO + OH \to CO_2 + H$ is known to be one of the plausible pathway for the formation of CO$_2$ ice \citep[e.g.,][]{Oba10,Iop11}. 
Numerical simulations of grain surface chemistry suggest that the efficiency of the CO$_2$ production via this pathway increases as the dust temperature increases, because the mobility of CO on the grain surface increases accordingly \citep[e.g.,][]{Ruf01,Cha12}. 
On the other hand, the formation of CH$_3$OH is believed to be mainly due to the hydrogenation of CO on grain surfaces \citep[e.g.,][]{Wat07} or photolysis/radiolysis of ice mantles \citep[e.g.,][]{Hud99,Ger01}. 
Astrochemical simulations suggest that the formation efficiency of CH$_3$OH via the CO hydrogenation rapidly decreases as the dust temperature increases \citep[e.g.,][]{Ruf01,Cup09,Cha12}. 
This is presumably due to the increased sublimation of hydrogen atoms and/or the consequence of the reaction-diffusion competition at elevated grain temperatures. 

Taking into account these theoretical and experimental implications, \citet{ST16} proposed the warm ice chemistry hypothesis to interpret the characteristic chemical compositions of ices in the LMC. 
The hypothesis argues that elevated dust temperatures in the LMC suppress the hydrogenation of CO, which leads to the low abundance of CH$_3$OH, while the warm dust increase the surface mobility of CO, leading to the enhanced production of CO$_2$. 
Astrochemical simulations dedicated to the environments of the LMC and SMC reproduce such a behavior \citep{Ach15,Ach16,Pau18}. 
Because solid CO is a highly volatile species (sublimation temperature for pure CO ice is $\sim$20 K), the non-detection of the CO ice in the SMC may be due to the even higher dust temperature in ice-forming dense clouds in the SMC as pointed out by \citet{Pau18}. 

It is likely that the ISM in low-metallicity star-forming galaxies has higher dust/gas temperature due to the reduced attenuation of the interstellar radiation field. 
For diffuse clouds, a good anti-correlation between metallicities and far-infrared color temperatures is reported for extragalaxies \citep[e.g.,][]{Eng08}. 
This would be interpreted as due to the increased interstellar radiation field strength in lower metallicity environments, which is caused by the reduced dust abundance. 
The dust temperature of ice-forming dense clouds in low-metallicity environments is, however, poorly understood so far. 
From the viewpoint of ice abundances discussed above, warmer dust is expected for a well-shielded ice-forming clouds in low-metallicity environments. 
Future direct temperature measurements of prestellar cores at low metallicity is highly required. 

As discussed in this section, the ice abundances in the LMC and SMC do not follow a simple metallicity scaling effect. 
It is important to take into account the indirect effect caused by the reduced dust abundance.

\section{Hot Molecular Cores at Low Metallicity}
Hot molecular cores are compact (size $<$0.1 pc), dense ($n_{\ce{H2}}$ $>$10$^6$ cm$^3$), and hot (T $>$100 K) protostellar sources, which appear during massive star formation. 
Hot cores are an excellent laboratory to study the molecular complexity of the ISM, since a variety of molecular species including COMs are often detected in solar-metallicity hot cores. 
The chemistry at this stage is characterized by the warm-up and sublimation of ice mantles that were formed in the earlier evolutionary stage, and also by the subsequent high-temperature gas-phase reactions. 

\begin{table}[btp!]
 \centering
 \caption{Summary of hot molecular cores detected in low-metallicity environments}\label{Tab_HC_Obj}
 {\tablefont\begin{tabular}{@{\extracolsep{\fill}} c c p{8em} c p{24em} }
    \midrule
    Galaxy & Metallicity (Z$_{\odot}$) & ALMA Band & No. & Reference  \\
    \midrule
    LMC & $\sim$0.4 & Band 6 (250 GHz), Band 7 (350 GHz) & 10 & \citet{ST16b, ST20}, Shimonishi et al. in prep., \citet{Sew18, Sew22a, Sew22b}, \citet{Gol24} \\
    SMC & $\sim$0.2 & Band 7 (350 GHz) & 2 & \citet{ST23} \\
    Outer MW & $\sim$0.3 & Band 6 (250 GHz), Band 7 (350 GHz) & 2 & \citet{ST21}, Ikeda et al. in prep.  \\
    \midrule
    \end{tabular}}
\tabnote{\textit{Notes}: All observations were carried out by ALMA. The third column ``No." indicates the total number of currently-known hot-core sources. }
\end{table}

Table \ref{Tab_HC_Obj} summarizes recent detection of hot cores in low-metallicity environments. 
The first detection of a low-metallicity (and also extragalactic) hot core is reported in \citet{ST16b} based on the observations of the high-mass protostar in the LMC with ALMA. 
Since then, hot cores have been detected in various low-metallicity environments of the LMC \citep[e.g.,][]{Sew18,ST20, Gol24}, the SMC \citep{ST23}, and the outer Galaxy \citep{ST21}, thanks to high-spatial-resolution and sensitive radio data brought by ALMA. 
Owing to the chemical richness of hot-core sources, various molecular species are now identified in low-metallicity environments as shown in Table \ref{Tab_HC_Mol}.

\begin{table}[tbp!]
 \centering
 \caption{Molecular species detected in low-metallicity hot cores}\label{Tab_HC_Mol}
 {\tablefont\begin{tabular}{@{\extracolsep{\fill}} p{6em} p{6em} p{5em} p{5em} p{5em} p{5em} p{5em} p{5em}}
    \midrule
    2 atoms & 3 atoms  & 4 atoms & 5 atoms & 6 atoms & 7 atoms  & 8 atoms & 9 atoms  \\
    \midrule
    \ce{CO} ($^{17}$O) & \ce{HCO+} ($^{13}$C,$^{18}$O,D) & \ce{H2CO} (D,\ce{D2}) & \ce{c-C3H2} & \ce{CH3OH} ($^{13}$C,D) & \ce{CH3CHO}   & \ce{HCOOCH3}  & \ce{CH3OCH3}  \\
    \ce{CN}            & \ce{HDO}                        & \ce{HNCO}             & \ce{HCOOH}  & \ce{CH3CN}              & \ce{c-C2H4O}  &               & \ce{C2H5OH}  \\
    \ce{NO}            & \ce{CCH}                        & \ce{H2CS}             & \ce{HC3N}   & \ce{NH2CHO}             &               &               & \ce{C2H5CN}  \\
    \ce{NS}            & \ce{HCN}                        &                       & \ce{H2CCO}  &                         &               &               &  \\
    \ce{CS} ($^{33,34}$S)  & \ce{H2S} (D)                &                       & & & & & \\
    \ce{SO} ($^{33,34}$S,$^{18,17}$O) & \ce{SO2} ($^{33,34}$S)                   & & & & & & \\
    \ce{SO+}           & \ce{OCS} ($^{13}$C)             &                       & & & & & \\
    \ce{SiO}           & & & & & & & \\
    \midrule
    \end{tabular}}
\tabnote{\textit{Notes}: The detected isotopologues are indicated in parenthesis. }
\end{table}

\begin{figure*}[ptbh!]
 \begin{center}
  \includegraphics[width=9.0cm]{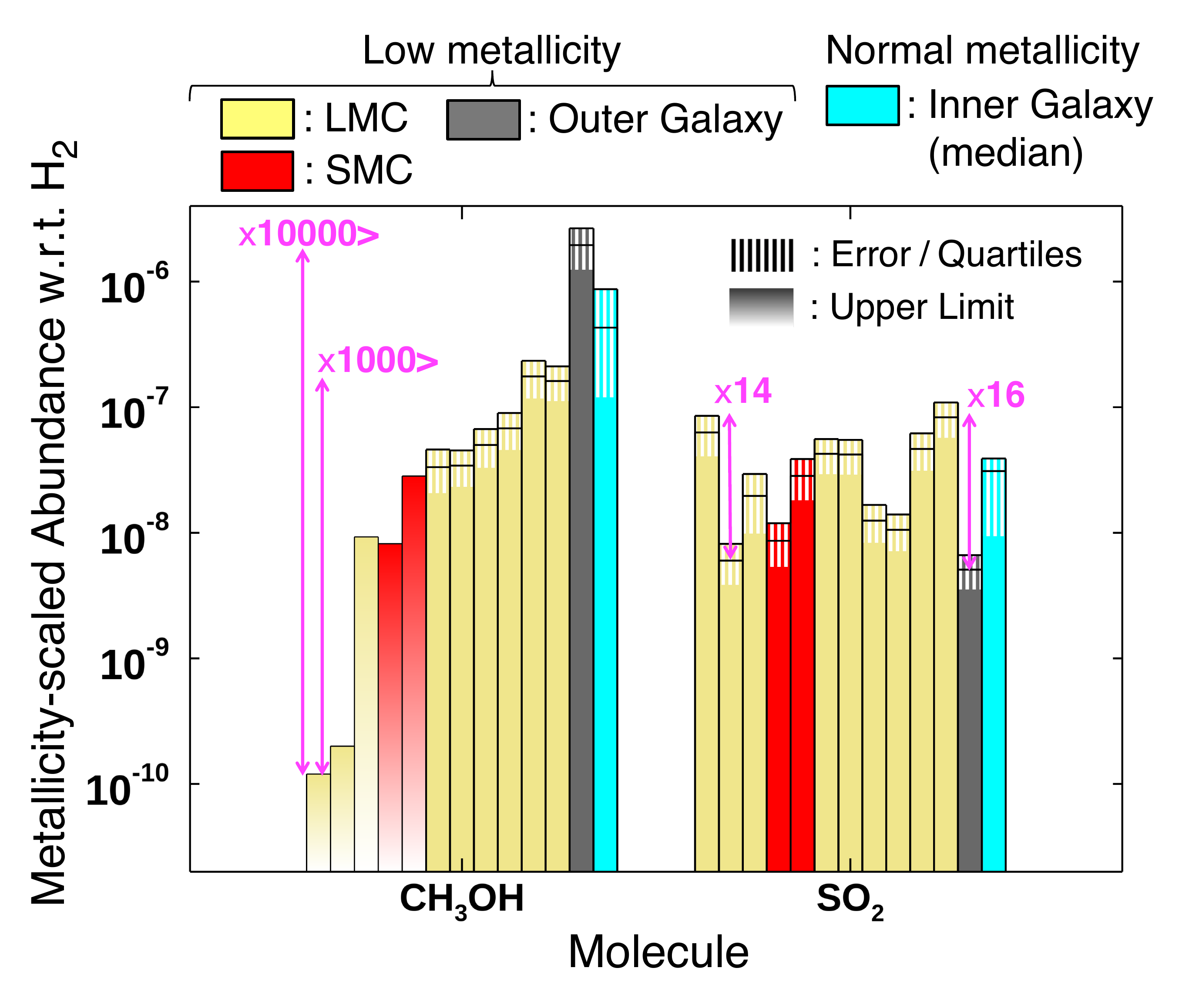}
  \caption{Comparison of metallicity-scaled molecular abundances of hot cores for LMC (yellow), SMC (red), outer Galaxy (gray), and inner Galaxy sources (cyan). 
  Abundances of low-metallicity hot cores are scaled in accordance with their metallicities to correct for the metallicity difference. 
  Only the contribution from high-temperature gas components (T $>$100 K) is plotted. 
  The area with thin vertical lines indicate the error bar for low-metallicity sources, while those of Galactic sources indicate the lower and upper quartiles of the abundance distribution. 
  The bar with a color gradient indicate an upper limit. 
  The plotted data are collected from the literature summarized in Table \ref{Tab_HC_Obj} for low-metallicity sources, while those for solar-metallicity sources are based on a sample of $\sim$30 inner Galaxy hot cores reported in \citet{Schi97, Hel97, Bae22}. 
  }
  \label{fig_HCAbu}
 \end{center}
\end{figure*}

Do the molecular abundances of low-metallicity hot cores simply scale with the metallicity?
Figure \ref{fig_HCAbu} shows a comparison of metallicity-scaled molecular abundances of high-temperature CH$_3$OH and SO$_2$ gas for hot cores in various metallicity environments. 
Both CH$_3$OH and SO$_2$ are believed to be tracer molecules of compact and high-temperature hot-core regions. 
All of the plotted sources exhibit high rotation temperatures above 100K in CH$_3$OH and/or SO$_2$ lines. 
Such high temperature is sufficient for the sublimation of ice mantles under the thermalized condition. 

\begin{figure*}[btp!]
 \begin{center}
  \includegraphics[width=9.5cm]{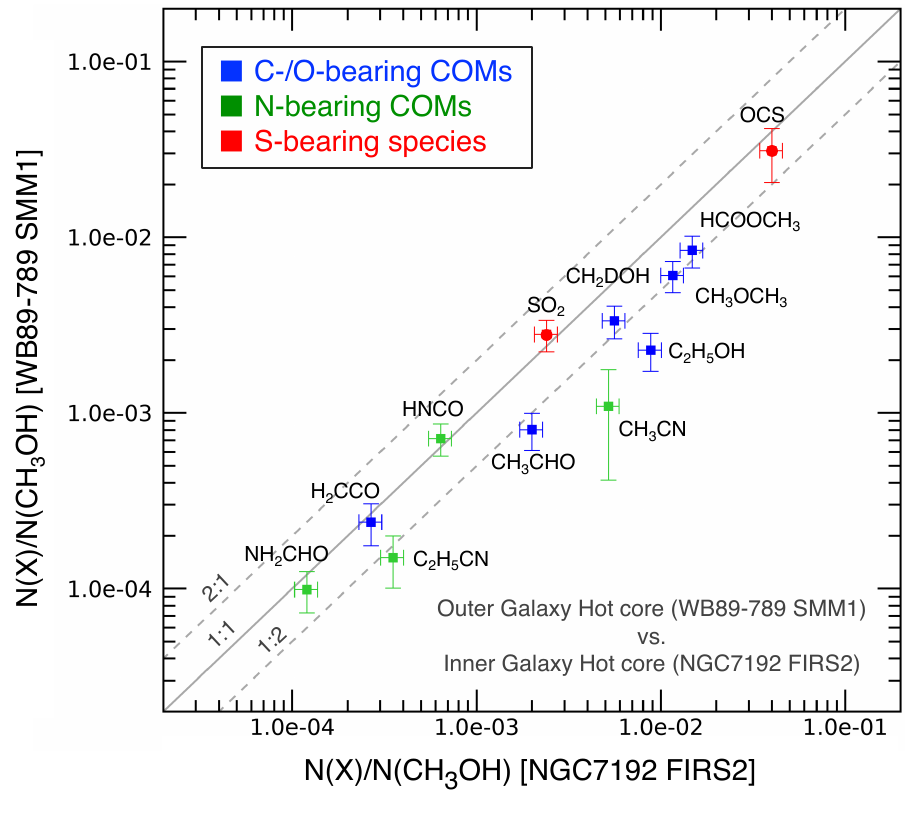}
  \caption{
  Comparison of molecular abundances normalized by the CH$_3$OH column density between a low-metallicity hot core (WB89-789 SMM1 in the extreme outer Galaxy) and a solar-metallicity hot core (NGC7192 FIRS2 in the inner Galaxy) The plotted data is based on \citet{ST21} for WB89-789 SMM1 and \citet{Fue14} for NGC7192 FIRS2, respectively. 
  Carbon- and oxygen-bearing species are shown by the blue squares, nitrogen-bearing species in green, and sulfur-bearing species in red. 
  The dotted lines represent an abundance ratio of 2:1 and 1:2 for WB89-789 SMM1 : NGC7192 FIRS2, while the solid line represent that of 1:1. 
  }
  \label{fig_EOGHCAbu}
 \end{center}
\end{figure*}

As seen in Figure \ref{fig_HCAbu}, the CH$_3$OH abundance shows large source-to-source variation even after corrected for the metallicity, suggesting that the chemical processes leading to the CH$_3$OH formation cannot be accounted for by a simple metallicity scaling effect. 
Such a large abundance variation is not seen in an inorganic molecule, SO$_2$, which is another commonly-used hot-core tracer. 

The reason for the large abundance diversity of an organic molecule, CH$_3$OH, in low-metallicity hot cores is still under debate. 
Several factors may potentially affect the derived COMs abundances. 
These include the variation in the source size/structure, the evolutionary stage, the degree of shock, the local physical conditions, etc. 
Since the plotted CH$_3$OH abundances are derived assuming the optically-thin emission, the optical thickness may have some effect, but it would not change the abundance by orders of magnitude. 

Astrochemical simulations for the chemical evolution of LMC/SMC hot cores suggest that dust temperature at the initial ice-forming stage has a significant effect on the CH$_3$OH gas abundance in the subsequent hot core stage, because the formation of solid CH$_3$OH is sensitive to dust temperature and it is inhibited on warmer grain surfaces \citep{Ach18,ST20}. 
Therefore, the diversity in the initial condition of star formation (e.g., degree of shielding, local radiation field strength) may lead to the large abundance variation of CH$_3$OH. 
As discussed in the previous section, infrared ice observations argue that the inhibition of the CH$_3$OH formation in the ice-forming stage is more likely to occur in the LMC condition due to the lower dust abundance and the stronger interstellar radiation field \citep{ST16}. 
Sensitive infrared observations of solid CH$_3$OH towards organic-rich and organic-poor hot cores in low-metallicity environments are highly awaited. 

\begin{figure*}[btp!]
 \begin{center}
  \includegraphics[width=13cm]{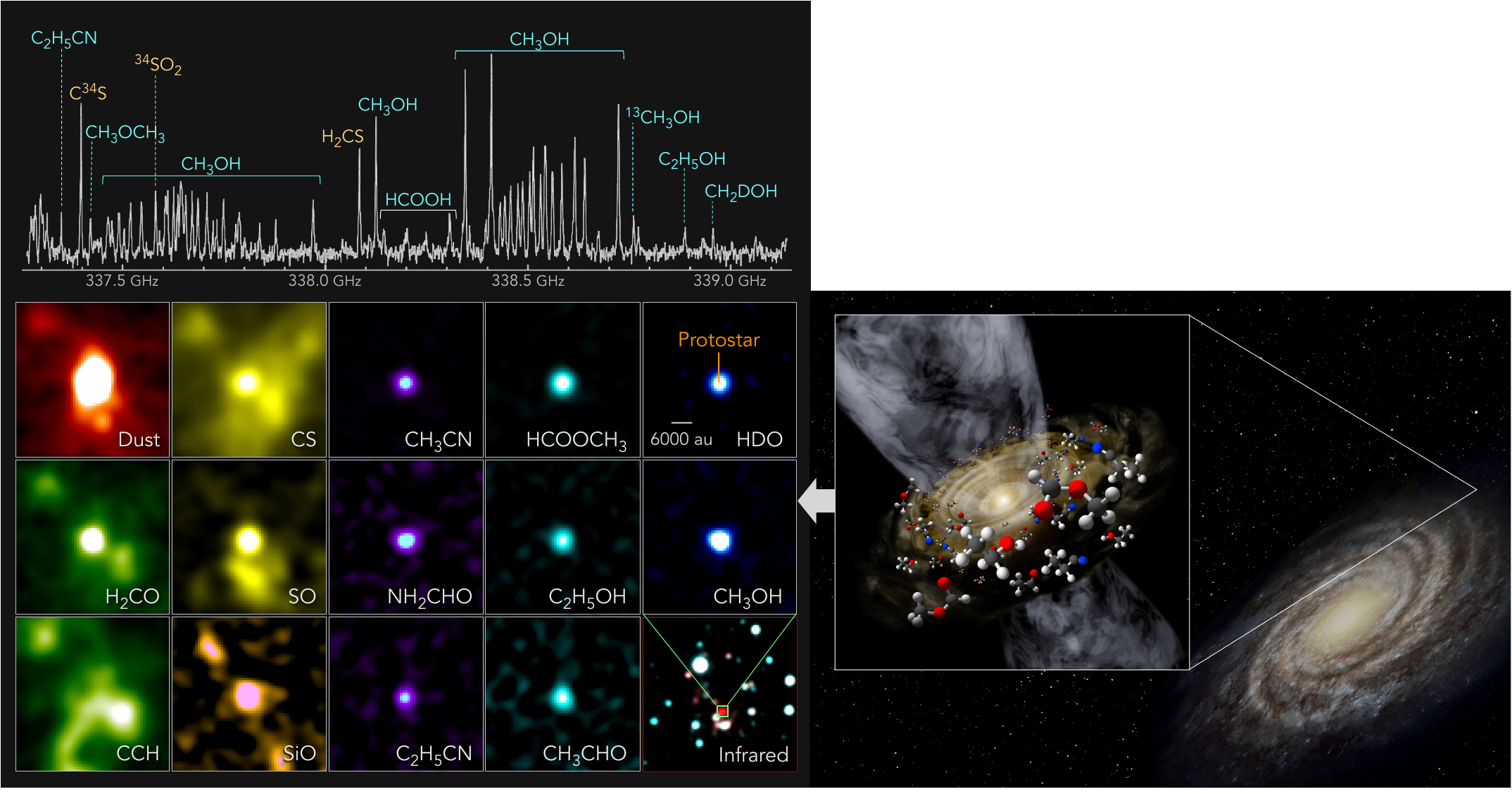}
  \caption{Molecular complexity discovered at the edge of our Galaxy. 
  Top left: Example of an ALMA submillimeter spectrum of the protostellar object, WB89-789 SMM1, found in the extreme outer Galaxy (galactocentric distance $\sim$19 kpc) \citep{ST21}. 
  Bottom left: Spatial distributions of dust and molecular line emission observed towards WB89-789 SMM1. 
  An infrared 2-color composite image of the surrounding region is also shown in the bottom right panel (red: 2.16 $\mu$m, blue: 1.25 $\mu$m, based on 2MASS) [Credit: ALMA (ESO/NAOJ/NRAO), T. Shimonishi]. 
  Right: Artist's conceptual image of the WB89-789 SMM1 protostar discovered in the extreme outer Galaxy [Credit: Niigata University]. 
  }
  \label{fig_EOGHC}
 \end{center}
\end{figure*}

What about COMs larger than CH$_3$OH?
Figure \ref{fig_EOGHCAbu} shows fractional abundances of COMs normalized by the CH$_3$OH column density. 
CH$_3$OH-normalized abundances of COMs are often discussed in protostellar cores since CH$_3$OH is believed to be a parental molecule for the formation of even larger COMs \citep[e.g.,][]{NM04,Gar06}. 
In the figure, the COMs data for the low-metallicity hot core found in the extreme outer Galaxy is compared with that of a solar-metallicity hot core. 
An example of the ALMA data for the former low-metallicity source is shown in Figure \ref{fig_EOGHC}.

Figure \ref{fig_EOGHCAbu} suggests that the $N$(X)/$N$(CH$_3$OH) ratios are remarkably similar between low-metallicity and solar-metallicity hot cores. 
This would imply that chemical processes leading to the formation of large COMs are regulated by the availability of CH$_3$OH, and once CH$_3$OH is provided, the COMs formation efficiency may not depend largely on the metallicity environment. 
We note that, however, the formation of CH$_3$OH itself is sometimes significantly inhibited in several low-metallicity hot cores in the LMC as shown in Figure \ref{fig_HCAbu}.
Future observations need to conduct systematic studies of large COMs towards protostellar cores located in various metallicity environments. 
Such studies will significantly impact our understanding of the molecular complexity in diverse interstellar environments.

\section{Summary and Outlook}
Metallicity is clearly one of the essential factors for the chemical evolution of star forming regions as discussed in this proceeding. 
Understanding the ISM chemistry as a function of metallicity will help glimpse the chemical processing in the early universe, and also it will help probe the diversity of the chemical evolution in various metallicity environments of the present-day Galaxy. 

The scaling of molecular abundances according to the elemental abundances of the natal environment is one of the important roles of the metallicity. 
However, it is also crucial to consider the indirect effect caused by the reduced dust abundance, which varies the radiation environment of star-forming regions. 

Recent detection of various molecules in low-metallicity star-forming regions suggests that great molecular complexity can be formed even in the reduced metallicity environments (1/2--1/10 Z$_{\odot}$), which is similar to the universe of 6--10 Gyr ago. 
In their chemical compositions, we clearly see the indirect effect caused by the reduced dust abundance. 
Enhanced photochemistry observed in low-metallicity molecular clouds would be the consequence of the reduced shielding of the interstellar radiation field. 
Ice compositions observed towards low-metallicity embedded protostars cannot be simply interpreted by the metallicity scaling effect, and we need to consider the grain surface chemistry on relatively warm dust compared to solar-metallicity counterparts. 
A variety of COMs are detected in low-metallicity hot cores with ALMA. 
Interestingly, the abundance of CH$_3$OH, one of the simplest COMs, shows a large source-to-source variation: it is roughly a metallicity-scaled abundance of Galactic counterparts in some sources, while it is significantly depleted beyond the level of metallicity difference in other sources. 
The reason of such a chemical diversity in low-metallicity protostellar sources is still under debate. 

Observations of the molecular complexity in nearby low-metallicity systems have made great and rapid progress in the last decade. 
As a future perspective, we still need to (i) increase the sample of protostellar cores in various metallicity environments (e.g., LMC/SMC/outer Galaxy as a low metallicity laboratory, while Galactic center region as a high metallicity counterpart), (ii) conduct radio and infrared follow-up observations towards currently reported low-metallicity protostellar cores with higher spatial resolution, higher sensitivity,  and/or wider spectral coverage, and (iii) refine astrochemical simulations of star-forming cores in various metallicity environments, with particular focus on the chemistry of organic molecules.

\section*{Acknowledgments}
I would like to thank the organizers for inviting me to the conference. 
This work was supported by JSPS KAKENHI grant Nos. JP20H05845, JP21H00037, and JP21H01145, and by Leading Initiative for Excellent Young Researchers, MEXT, Japan. 
I would like to thank organizers for inviting me to the Kavli-IAU Astrochemistry Symposium.


\begin{thebibliography}{}

\bibitem[{Acharyya} and {Herbst}, 2015]{Ach15}
{Acharyya}, K. \& {Herbst}, E. 2015, {Molecular Development in the Large Magellanic Cloud}.
\newblock {\em \apj}, 812, 142.

\bibitem[{Acharyya} and {Herbst}, 2016]{Ach16}
{Acharyya}, K. \& {Herbst}, E. 2016, {Simulations of the Chemistry in the Small Magellanic Cloud}.
\newblock {\em \apj}, 822, 105.

\bibitem[{Acharyya} and {Herbst}, 2018]{Ach18}
{Acharyya}, K. \& {Herbst}, E. 2018, {Hot Cores in Magellanic Clouds}.
\newblock {\em \apj}, 859, 51.

\bibitem[{An} et~al., 2011]{An11}
{An}, D., {Ram{\'\i}rez}, S.~V., {Sellgren}, K., {Arendt}, R.~G., {Adwin Boogert}, A.~C., {Robitaille}, T.~P., {Schultheis}, M., {Cotera}, A.~S., {Smith}, H.~A., \& {Stolovy}, S.~R. 2011, {Massive Young Stellar Objects in the Galactic Center. I. Spectroscopic Identification from Spitzer Infrared Spectrograph Observations}.
\newblock {\em \apj}, 736(2), 133.

\bibitem[{An} et~al., 2009]{An09}
{An}, D., {Ram{\'\i}rez}, S.~V., {Sellgren}, K., {Arendt}, R.~G., {Boogert}, A.~C.~A., {Schultheis}, M., {Stolovy}, S.~R., {Cotera}, A.~S., {Robitaille}, T.~P., \& {Smith}, H.~A. 2009, {First Spectroscopic Identification of Massive Young Stellar Objects in the Galactic Center}.
\newblock {\em \apjl}, 702(2), L128--L132.

\bibitem[{Arellano-C{\'o}rdova} et~al., 2020]{Are20}
{Arellano-C{\'o}rdova}, K.~Z., {Esteban}, C., {Garc{\'\i}a-Rojas}, J., \& {M{\'e}ndez-Delgado}, J.~E. 2020, {The Galactic radial abundance gradients of C, N, O, Ne, S, Cl, and Ar from deep spectra of H II regions}.
\newblock {\em \mnras}, 496(2), 1051--1076.

\bibitem[{Baek} et~al., 2022]{Bae22}
{Baek}, G., {Lee}, J.-E., {Hirota}, T., {Kim}, K.-T., \& {Kim}, M.~K. 2022, {Complex Organic Molecules Detected in 12 High-mass Star-forming Regions with Atacama Large Millimeter/submillimeter Array}.
\newblock {\em \apj}, 939(2), 84.

\bibitem[{Balestra} et~al., 2007]{Bal07}
{Balestra}, I., {Tozzi}, P., {Ettori}, S., {Rosati}, P., {Borgani}, S., {Mainieri}, V., {Norman}, C., \& {Viola}, M. 2007, {Tracing the evolution in the iron content of the intra-cluster medium}.
\newblock {\em \aap}, 462(2), 429--442.

\bibitem[{Bernal} et~al., 2021]{Ber21}
{Bernal}, J.~J., {Sephus}, C.~D., \& {Ziurys}, L.~M. 2021, {Methanol at the Edge of the Galaxy: New Observations to Constrain the Galactic Habitable Zone}.
\newblock {\em \apj}, 922(2), 106.

\bibitem[{Blair} et~al., 2008]{Bla08}
{Blair}, S.~K., {Magnani}, L., {Brand}, J., \& {Wouterloot}, J. G.~A. 2008, {Formaldehyde in the Far Outer Galaxy: Constraining the Outer Boundary of the Galactic Habitable Zone}.
\newblock {\em Astrobiology}, 8(1), 59--73.

\bibitem[{Boogert} et~al., 2015]{Boo15}
{Boogert}, A., {Gerakines}, P., \& {Whittet}, D. 2015, {Observations of the Icy Universe}.
\newblock {\em ArXiv e-prints},.

\bibitem[{Braine} et~al., 2017]{Bra17}
{Braine}, J., {Shimajiri}, Y., {Andr{\'e}}, P., {Bontemps}, S., {Gao}, Y., {Chen}, H., \& {Kramer}, C. 2017, {Dense gas in low-metallicity galaxies}.
\newblock {\em \aap}, 597, A44.

\bibitem[{Braine} et~al., 2023]{Bra23}
{Braine}, J., {Sun}, Y., {Shimajiri}, Y., {van der Tak}, F.~F.~S., {Fang}, M., {Andr{\'e}}, P., {Chen}, H., \& {Gao}, Y. 2023, {Dense gas and star formation in the outer Milky Way}.
\newblock {\em \aap}, 676, A27.

\bibitem[{Brand} and {Wouterloot}, 2007]{Bra07}
{Brand}, J. \& {Wouterloot}, J.~G.~A. 2007, {A star cluster at the edge of the Galaxy}.
\newblock {\em \aap}, 464(3), 909--920.

\bibitem[{Buchbender} et~al., 2013]{Buc13}
{Buchbender}, C., {Kramer}, C., {Gonzalez-Garcia}, M., {Israel}, F.~P., {Garc{\'\i}a-Burillo}, S., {van der Werf}, P., {Braine}, J., {Rosolowsky}, E., {Mookerjea}, B., {Aalto}, S., {Boquien}, M., {Gratier}, P., {Henkel}, C., {Quintana-Lacaci}, G., {Verley}, S., \& {van der Tak}, F. 2013, {Dense gas in M 33 (HerM33es)}.
\newblock {\em \aap}, 549, A17.

\bibitem[{Chang} and {Herbst}, 2012]{Cha12}
{Chang}, Q. \& {Herbst}, E. 2012, {A Unified Microscopic-Macroscopic Monte Carlo Simulation of Gas-grain Chemistry in Cold Dense Interstellar Clouds}.
\newblock {\em \apj}, 759, 147.

\bibitem[{Chiar} et~al., 2002]{Chi02}
{Chiar}, J.~E., {Adamson}, A.~J., {Pendleton}, Y.~J., {Whittet}, D.~C.~B., {Caldwell}, D.~A., \& {Gibb}, E.~L. 2002, {Hydrocarbons, Ices, and ``XCN'' in the Line of Sight toward the Galactic Center}.
\newblock {\em \apj}, 570(1), 198--209.

\bibitem[{Chiar} et~al., 2000]{Chi00}
{Chiar}, J.~E., {Tielens}, A.~G.~G.~M., {Whittet}, D.~C.~B., {Schutte}, W.~A., {Boogert}, A.~C.~A., {Lutz}, D., {van Dishoeck}, E.~F., \& {Bernstein}, M.~P. 2000, {The Composition and Distribution of Dust along the Line of Sight toward the Galactic Center}.
\newblock {\em \apj}, 537(2), 749--762.

\bibitem[{Chin} et~al., 1998]{Chin98}
{Chin}, Y.-N., {Henkel}, C., {Millar}, T.~J., {Whiteoak}, J.~B., \& {Marx-Zimmer}, M. 1998, {Molecular abundances in the Magellanic Clouds. III. LIRS36, a star-forming region in the Small Magellanic Cloud}.
\newblock {\em \aap}, 330, 901--909.

\bibitem[{Chin} et~al., 1996]{Chin96}
{Chin}, Y.-N., {Henkel}, C., {Millar}, T.~J., {Whiteoak}, J.~B., \& {Mauersberger}, R. 1996, {Molecular abundances in the Magellanic Clouds. II. Deuterated species in the LMC.}
\newblock {\em \aap}, 312, L33--L36.

\bibitem[{Chin} et~al., 1997]{Chin97}
{Chin}, Y.-N., {Henkel}, C., {Whiteoak}, J.~B., {Millar}, T.~J., {Hunt}, M.~R., \& {Lemme}, C. 1997, {Molecular abundances in the Magellanic Clouds. I. A multiline study of five cloud cores.}
\newblock {\em \aap}, 317, 548--562.

\bibitem[{Colzi} et~al., 2022]{Col22}
{Colzi}, L., {Romano}, D., {Fontani}, F., {Rivilla}, V.~M., {Bizzocchi}, L., {Beltran}, M.~T., {Caselli}, P., {Elia}, D., \& {Magrini}, L. 2022, {CHEMOUT: CHEMical complexity in star-forming regions of the OUTer Galaxy. III. Nitrogen isotopic ratios in the outer Galaxy}.
\newblock {\em \aap}, 667, A151.

\bibitem[{Cuppen} et~al., 2009]{Cup09}
{Cuppen}, H.~M., {van Dishoeck}, E.~F., {Herbst}, E., \& {Tielens}, A.~G.~G.~M. 2009, {Microscopic simulation of methanol and formaldehyde ice formation in cold dense cores}.
\newblock {\em \aap}, 508, 275--287.

\bibitem[{Dartois} et~al., 2002]{Dar02}
{Dartois}, E., {d'Hendecourt}, L., {Thi}, W., {Pontoppidan}, K.~M., \& {van Dishoeck}, E.~F. 2002, {Combined VLT ISAAC/ISO SWS spectroscopy of two protostellar sources. The importance of minor solid state features}.
\newblock {\em \aap}, 394, 1057--1068.

\bibitem[{Delgado Mena} et~al., 2019]{Del19}
{Delgado Mena}, E., {Moya}, A., {Adibekyan}, V., {Tsantaki}, M., {Gonz{\'a}lez Hern{\'a}ndez}, J.~I., {Israelian}, G., {Davies}, G.~R., {Chaplin}, W.~J., {Sousa}, S.~G., {Ferreira}, A.~C.~S., \& {Santos}, N.~C. 2019, {Abundance to age ratios in the HARPS-GTO sample with Gaia DR2. Chemical clocks for a range of [Fe/H]}.
\newblock {\em \aap}, 624, A78.

\bibitem[{Engelbracht} et~al., 2008]{Eng08}
{Engelbracht}, C.~W., {Rieke}, G.~H., {Gordon}, K.~D., {Smith}, J.-D.~T., {Werner}, M.~W., {Moustakas}, J., {Willmer}, C.~N.~A., \& {Vanzi}, L. 2008, {Metallicity Effects on Dust Properties in Starbursting Galaxies}.
\newblock {\em \apj}, 678, 804--827.

\bibitem[{Fern{\'a}ndez-Mart{\'\i}n} et~al., 2017]{Fer17}
{Fern{\'a}ndez-Mart{\'\i}n}, A., {P{\'e}rez-Montero}, E., {V{\'\i}lchez}, J.~M., \& {Mampaso}, A. 2017, {Chemical distribution of H II regions towards the Galactic anticentre}.
\newblock {\em \aap}, 597, A84.

\bibitem[{Fontani} et~al., 2022a]{Fon22a}
{Fontani}, F., {Colzi}, L., {Bizzocchi}, L., {Rivilla}, V.~M., {Elia}, D., {Beltr{\'a}n}, M.~T., {Caselli}, P., {Magrini}, L., {S{\'a}nchez-Monge}, A., {Testi}, L., \& {Romano}, D. 2022,a {CHEMOUT: CHEMical complexity in star-forming regions of the OUTer Galaxy. I. Organic molecules and tracers of star-formation activity}.
\newblock {\em \aap}, 660a, A76.

\bibitem[{Fontani} et~al., 2022b]{Fon22b}
{Fontani}, F., {Schmiedeke}, A., {S{\'a}nchez-Monge}, A., {Colzi}, L., {Elia}, D., {Rivilla}, V.~M., {Beltr{\'a}n}, M.~T., {Bizzocchi}, L., {Caselli}, P., {Magrini}, L., \& {Romano}, D. 2022,b {CHEMOUT: CHEMical complexity in star-forming regions of the OUTer Galaxy. II. Methanol formation at low metallicity}.
\newblock {\em \aap}, 664b, A154.

\bibitem[{Fuente} et~al., 2014]{Fue14}
{Fuente}, A., {Cernicharo}, J., {Caselli}, P., {McCoey}, C., {Johnstone}, D., {Fich}, M., {van Kempen}, T., {Palau}, A., {Y{\i}ld{\i}z}, U.~A., {Tercero}, B., \& {L{\'o}pez}, A. 2014, {The hot core towards the intermediate-mass protostar NGC 7129 FIRS 2. Chemical similarities with Orion KL}.
\newblock {\em \aap}, 568, A65.

\bibitem[{Galametz} et~al., 2020]{Gal20}
{Galametz}, M., {Schruba}, A., {De Breuck}, C., {Immer}, K., {Chevance}, M., {Galliano}, F., {Gusdorf}, A., {Lebouteiller}, V., {Lee}, M.~Y., {Madden}, S.~C., {Polles}, F.~L., \& {van Kempen}, T.~A. 2020, {DeGaS-MC: Dense Gas Survey in the Magellanic Clouds. I. An APEX survey of HCO$^{+}$ and HCN(2-1) toward the LMC and SMC}.
\newblock {\em \aap}, 643, A63.

\bibitem[{Garrod} and {Herbst}, 2006]{Gar06}
{Garrod}, R.~T. \& {Herbst}, E. 2006, {Formation of methyl formate and other organic species in the warm-up phase of hot molecular cores}.
\newblock {\em \aap}, 457, 927--936.

\bibitem[{Gerakines} et~al., 2001]{Ger01}
{Gerakines}, P.~A., {Moore}, M.~H., \& {Hudson}, R.~L. 2001, {Energetic processing of laboratory ice analogs: UV photolysis versus ion bombardment}.
\newblock {\em \jgr}, 106(E12), 33381--33386.

\bibitem[{Gibb} et~al., 2004]{Gib04}
{Gibb}, E.~L., {Whittet}, D.~C.~B., {Boogert}, A.~C.~A., \& {Tielens}, A.~G.~G.~M. 2004, {Interstellar Ice: The Infrared Space Observatory Legacy}.
\newblock {\em \apjs}, 151, 35--73.

\bibitem[{Gibb} et~al., 2001]{Gib01}
{Gibb}, E.~L., {Whittet}, D.~C.~B., \& {Chiar}, J.~E. 2001, {Searching for Ammonia in Grain Mantles toward Massive Young Stellar Objects}.
\newblock {\em \apj}, 558, 702--716.

\bibitem[{Gong} et~al., 2023]{Gon23}
{Gong}, Y., {Henkel}, C., {Menten}, K.~M., {Chen}, C. H.~R., {Zhang}, Z.~Y., {Yan}, Y.~T., {Weiss}, A., {Langer}, N., {Wang}, J.~Z., {Mao}, R.~Q., {Tang}, X.~D., {Yang}, W., {Ao}, Y.~P., \& {Wang}, M. 2023, {Sulfur isotope ratios in the Large Magellanic Cloud}.
\newblock {\em \aap}, 679, L6.

\bibitem[{Graczyk} et~al., 2014]{Gra14}
{Graczyk}, D., {Pietrzy{\'n}ski}, G., {Thompson}, I.~B., {Gieren}, W., {Pilecki}, B., {Konorski}, P., {Udalski}, A., {Soszy{\'n}ski}, I., {Villanova}, S., {G{\'o}rski}, M., {Suchomska}, K., {Karczmarek}, P., {Kudritzki}, R.-P., {Bresolin}, F., \& {Gallenne}, A. 2014, {The Araucaria Project. The Distance to the Small Magellanic Cloud from Late-type Eclipsing Binaries}.
\newblock {\em \apj}, 780, 59.

\bibitem[{Hamedani Golshan} et~al., 2024]{Gol24}
{Hamedani Golshan}, R., {S{\'a}nchez-Monge}, {\'A}., {Schilke}, P., {Sewi{\l}o}, M., {M{\"o}ller}, T., {Veena}, V.~S., \& {Fuller}, G.~A. 2024, {High-mass star formation across the Large Magellanic Cloud I. Chemical properties and hot molecular cores observed with ALMA at 1.2 mm}.
\newblock {\em arXiv e-prints},, arXiv:2405.01710.

\bibitem[{Heikkil{\"a}} et~al., 1997]{Hei97}
{Heikkil{\"a}}, A., {Johansson}, L.~E.~B., \& {Olofsson}, H. 1997, {Observation of DCO\^+\^ in the Large Magellanic Cloud.}
\newblock {\em \aap}, 319, L21--L24.

\bibitem[{Heikkil{\"a}} et~al., 1999]{Hei99}
{Heikkil{\"a}}, A., {Johansson}, L.~E.~B., \& {Olofsson}, H. 1999, {Molecular abundance variations in the Magellanic Clouds}.
\newblock {\em \aap}, 344, 817--847.

\bibitem[{Helmich} and {van Dishoeck}, 1997]{Hel97}
{Helmich}, F.~P. \& {van Dishoeck}, E.~F. 1997, {Physical and chemical variations within the W3 star-forming region. II. The 345 GHz spectral line survey.}
\newblock {\em \aaps}, 124.

\bibitem[{Herbst} and {van Dishoeck}, 2009]{Her09}
{Herbst}, E. \& {van Dishoeck}, E.~F. 2009, {Complex Organic Interstellar Molecules}.
\newblock {\em \araa}, 47, 427--480.

\bibitem[{Houck} et~al., 2004]{Hou04}
{Houck}, J.~R., {Roellig}, T.~L., {van Cleve}, J., {Forrest}, W.~J., {Herter}, T., {Lawrence}, C.~R., {Matthews}, K., {Reitsema}, H.~J., {Soifer}, B.~T., {Watson}, D.~M., {Weedman}, D., {Huisjen}, M., {Troeltzsch}, J., {Barry}, D.~J., {Bernard-Salas}, J., {Blacken}, C.~E., {Brandl}, B.~R., {Charmandaris}, V., {Devost}, D., {Gull}, G.~E., {Hall}, P., {Henderson}, C.~P., {Higdon}, S.~J.~U., {Pirger}, B.~E., {Schoenwald}, J., {Sloan}, G.~C., {Uchida}, K.~I., {Appleton}, P.~N., {Armus}, L., {Burgdorf}, M.~J., {Fajardo-Acosta}, S.~B., {Grillmair}, C.~J., {Ingalls}, J.~G., {Morris}, P.~W., \& {Teplitz}, H.~I. 2004, {The Infrared Spectrograph (IRS) on the Spitzer Space Telescope}.
\newblock {\em \apjs}, 154, 18--24.

\bibitem[{Hudson} and {Moore}, 1999]{Hud99}
{Hudson}, R.~L. \& {Moore}, M.~H. 1999, {Laboratory Studies of the Formation of Methanol and Other Organic Molecules by Water+Carbon Monoxide Radiolysis: Relevance to Comets, Icy Satellites, and Interstellar Ices}.
\newblock {\em \icarus}, 140(2), 451--461.

\bibitem[{Hunter} et~al., 2005]{Hun05}
{Hunter}, I., {Dufton}, P.~L., {Ryans}, R.~S.~I., {Lennon}, D.~J., {Rolleston}, W.~R.~J., {Hubeny}, I., \& {Lanz}, T. 2005, {A non-LTE analysis of the spectra of two narrow lined main sequence stars in the SMC}.
\newblock {\em \aap}, 436(2), 687--695.

\bibitem[{Ioppolo} et~al., 2011]{Iop11}
{Ioppolo}, S., {van Boheemen}, Y., {Cuppen}, H.~M., {van Dishoeck}, E.~F., \& {Linnartz}, H. 2011, {Surface formation of CO$_{2}$ ice at low temperatures}.
\newblock {\em \mnras}, 413, 2281--2287.

\bibitem[{Johansson}, 1991]{Joh91}
{Johansson}, L.~E.~B.
\newblock {Interstellar Gas in the Magellanic Clouds: SEST Observations of CO and Other Molecules}.
\newblock In {Combes}, F. \& {Casoli}, F., editors, {\em Dynamics of Galaxies and Their Molecular Cloud Distributions} 1991,, volume 146, ~1.

\bibitem[{Johansson} et~al., 1994]{Joh94}
{Johansson}, L.~E.~B., {Olofsson}, H., {Hjalmarson}, A., {Gredel}, R., \& {Black}, J.~H. 1994, {Interstellar molecules in the Large Magellanic Cloud}.
\newblock {\em \aap}, 291, 89--105.

\bibitem[{Kepley} et~al., 2018]{Kep18}
{Kepley}, A.~A., {Bittle}, L., {Leroy}, A.~K., {Jim{\'e}nez-Donaire}, M.~J., {Schruba}, A., {Bigiel}, F., {Gallagher}, M., {Johnson}, K., \& {Usero}, A. 2018, {Dense Molecular Gas in the Nearby Low-metallicity Dwarf Starburst Galaxy IC 10}.
\newblock {\em \apj}, 862(2), 120.

\bibitem[{Madau} and {Dickinson}, 2014]{Mad14}
{Madau}, P. \& {Dickinson}, M. 2014, {Cosmic Star-Formation History}.
\newblock {\em \araa}, 52, 415--486.

\bibitem[{Madden} and {Cormier}, 2019]{Mad19}
{Madden}, S.~C. \& {Cormier}, D.
\newblock {Dwarf Galaxies: Their Low Metallicity Interstellar Medium}.
\newblock In {McQuinn}, K. B.~W. \& {Stierwalt}, S., editors, {\em Dwarf Galaxies: From the Deep Universe to the Present} 2019,, volume 344, pp. 240--254.

\bibitem[{Magrini} and {Gon{\c{c}}alves}, 2009]{Mag09a}
{Magrini}, L. \& {Gon{\c{c}}alves}, D.~R. 2009, {IC10: the history of the nearest starburst galaxy through its Planetary Nebula and HII region populations}.
\newblock {\em \mnras}, 398(1), 280--292.

\bibitem[{Magrini} et~al., 2009]{Mag09b}
{Magrini}, L., {Stanghellini}, L., \& {Villaver}, E. 2009, {The Planetary Nebula Population of M33 and its Metallicity Gradient: A Look Into the Galaxy's Distant Past}.
\newblock {\em \apj}, 696(1), 729--740.

\bibitem[{Mart{\'\i}n} et~al., 2014]{Mar14}
{Mart{\'\i}n}, S., {Verdes-Montenegro}, L., {Aladro}, R., {Espada}, D., {Argudo-Fern{\'a}ndez}, M., {Kramer}, C., \& {Scott}, T.~C. 2014, {Chemistry in isolation: High CCH/HCO$^{+}$ line ratio in the AMIGA galaxy CIG 638}.
\newblock {\em \aap}, 563, L6.

\bibitem[{Moorwood} et~al., 1998]{Moo98}
{Moorwood}, A., {Cuby}, J.~G., {Biereichel}, P., {Brynnel}, J., {Delabre}, B., {Devillard}, N., {van Dijsseldonk}, A., {Finger}, G., {Gemperlein}, H., {Gilmozzi}, R., {Herlin}, T., {Huster}, G., {Knudstrup}, J., {Lidman}, C., {Lizon}, J.~L., {Mehrgan}, H., {Meyer}, M., {Nicolini}, G., {Petr}, M., {Spyromilio}, J., \& {Stegmeier}, J. 1998, {ISAAC sees first light at the VLT.}
\newblock {\em The Messenger}, 94, 7--9.

\bibitem[{Murakami} et~al., 2007]{Mur07}
{Murakami}, H., {Baba}, H., {Barthel}, P., {Clements}, D.~L., {Cohen}, M., {Doi}, Y., {Enya}, K., {Figueredo}, E., {Fujishiro}, N., {Fujiwara}, H., {Fujiwara}, M., {Garcia-Lario}, P., {Goto}, T., {Hasegawa}, S., {Hibi}, Y., {Hirao}, T., {Hiromoto}, N., {Hong}, S.~S., {Imai}, K., {Ishigaki}, M., {Ishiguro}, M., {Ishihara}, D., {Ita}, Y., {Jeong}, W., {Jeong}, K.~S., {Kaneda}, H., {Kataza}, H., {Kawada}, M., {Kawai}, T., {Kawamura}, A., {Kessler}, M.~F., {Kester}, D., {Kii}, T., {Kim}, D.~C., {Kim}, W., {Kobayashi}, H., {Koo}, B.~C., {Kwon}, S.~M., {Lee}, H.~M., {Lorente}, R., {Makiuti}, S., {Matsuhara}, H., {Matsumoto}, T., {Matsuo}, H., {Matsuura}, S., {M{\"u}ller}, T.~G., {Murakami}, N., {Nagata}, H., {Nakagawa}, T., {Naoi}, T., {Narita}, M., {Noda}, M., {Oh}, S.~H., {Ohnishi}, A., {Ohyama}, Y., {Okada}, Y., {Okuda}, H., {Oliver}, S., {Onaka}, T., {Ootsubo}, T., {Oyabu}, S., {Pak}, S., {Park}, Y., {Pearson}, C.~P., {Rowan-Robinson}, M., {Saito}, T., {Sakon}, I., {Salama}, A., {Sato}, S., {Savage}, R.~S.,
  {Serjeant}, S., {Shibai}, H., {Shirahata}, M., {Sohn}, J., {Suzuki}, T., {Takagi}, T., {Takahashi}, H., {Tanab{\'e}}, T., {Takeuchi}, T.~T., {Takita}, S., {Thomson}, M., {Uemizu}, K., {Ueno}, M., {Usui}, F., {Verdugo}, E., {Wada}, T., {Wang}, L., {Watabe}, T., {Watarai}, H., {White}, G.~J., {Yamamura}, I., {Yamauchi}, C., \& {Yasuda}, A. 2007, {The Infrared Astronomical Mission AKARI}.
\newblock {\em \pasj}, 59, 369--+.

\bibitem[{Nishimura} et~al., 2016b]{Nis16b}
{Nishimura}, Y., {Shimonishi}, T., {Watanabe}, Y., {Sakai}, N., {Aikawa}, Y., {Kawamura}, A., \& {Yamamoto}, S. 2016,b {Spectral Line Survey toward a Molecular Cloud in IC10}.
\newblock {\em \apj}, 829b, 94.

\bibitem[{Nishimura} et~al., 2016a]{Nis16a}
{Nishimura}, Y., {Shimonishi}, T., {Watanabe}, Y., {Sakai}, N., {Aikawa}, Y., {Kawamura}, A., \& {Yamamoto}, S. 2016,a {Spectral Line Survey toward Molecular Clouds in the Large Magellanic Cloud}.
\newblock {\em \apj}, 818a, 161.

\bibitem[{Nomura} and {Millar}, 2004]{NM04}
{Nomura}, H. \& {Millar}, T.~J. 2004, {The physical and chemical structure of hot molecular cores}.
\newblock {\em \aap}, 414, 409--423.

\bibitem[{Oba} et~al., 2010]{Oba10}
{Oba}, Y., {Watanabe}, N., {Kouchi}, A., {Hama}, T., \& {Pirronello}, V. 2010, {Experimental Study of CO$_{2}$ Formation by Surface Reactions of Non-energetic OH Radicals with CO Molecules}.
\newblock {\em \apjl}, 712, L174--L178.

\bibitem[{Oliveira} et~al., 2009]{Oli09}
{Oliveira}, J.~M., {van Loon}, J.~T., {Chen}, C., {Tielens}, A.~G.~G.~M., {Sloan}, G.~C., {Woods}, P.~M., {Kemper}, F., {Indebetouw}, R., {Gordon}, K.~D., {Boyer}, M.~L., {Shiao}, B., {Madden}, S., {Speck}, A.~K., {Meixner}, M., \& {Marengo}, M. 2009, {Ice Chemistry in Embedded Young Stellar Objects in the Large Magellanic Cloud}.
\newblock {\em \apj}, 707, 1269--1295.

\bibitem[{Oliveira} et~al., 2011]{Oli11}
{Oliveira}, J.~M., {van Loon}, J.~T., {Sloan}, G.~C., {Indebetouw}, R., {Kemper}, F., {Tielens}, A.~G.~G.~M., {Simon}, J.~D., {Woods}, P.~M., \& {Meixner}, M. 2011, {Ice chemistry in massive young stellar objects: the role of metallicity}.
\newblock {\em \mnras}, 411, L36--L40.

\bibitem[{Oliveira} et~al., 2013]{Oli13}
{Oliveira}, J.~M., {van Loon}, J.~T., {Sloan}, G.~C., {Sewi{\l}o}, M., {Kraemer}, K.~E., {Wood}, P.~R., {Indebetouw}, R., {Filipovi{\'c}}, M.~D., {Crawford}, E.~J., {Wong}, G.~F., {Hora}, J.~L., {Meixner}, M., {Robitaille}, T.~P., {Shiao}, B., \& {Simon}, J.~D. 2013, {Early-stage young stellar objects in the Small Magellanic Cloud}.
\newblock {\em \mnras}, 428, 3001--3033.

\bibitem[{Oliveira} et~al., 2006]{Oli06}
{Oliveira}, J.~M., {van Loon}, J.~T., {Stanimirovi{\'c}}, S., \& {Zijlstra}, A.~A. 2006, {Massive young stellar objects in the Large Magellanic Cloud: water masers and ESO-VLT 3-4 {$\mu$}m spectroscopy}.
\newblock {\em \mnras}, 372, 1509--1524.

\bibitem[{Onaka} et~al., 2007]{TON07}
{Onaka}, T., {Matsuhara}, H., {Wada}, T., {Fujishiro}, N., {Fujiwara}, H., {Ishigaki}, M., {Ishihara}, D., {Ita}, Y., {Kataza}, H., {Kim}, W., {Matsumoto}, T., {Murakami}, H., {Ohyama}, Y., {Oyabu}, S., {Sakon}, I., {Tanab{\'e}}, T., {Takagi}, T., {Uemizu}, K., {Ueno}, M., {Usui}, F., {Watarai}, H., {Cohen}, M., {Enya}, K., {Ootsubo}, T., {Pearson}, C.~P., {Takeyama}, N., {Yamamuro}, T., \& {Ikeda}, Y. 2007, {The Infrared Camera (IRC) for AKARI -- Design and Imaging Performance}.
\newblock {\em \pasj}, 59, 401--+.

\bibitem[{Paron} et~al., 2014]{Par14}
{Paron}, S., {Ortega}, M.~E., {Cunningham}, M., {Jones}, P.~A., {Rubio}, M., {Fari{\~n}a}, C., \& {Komugi}, S. 2014, {ASTE observations in the 345 GHz window towards the HII region N113 of the Large Magellanic Cloud}.
\newblock {\em \aap}, 572, A56.

\bibitem[{Paron} et~al., 2016]{Par16}
{Paron}, S., {Ortega}, M.~E., {Fari{\~n}a}, C., {Cunningham}, M., {Jones}, P.~A., \& {Rubio}, M. 2016, {A view of Large Magellanic Cloud H II regions N159, N132, and N166 through the 345-GHz window}.
\newblock {\em \mnras}, 455, 518--525.

\bibitem[{Patra} et~al., 2022]{Pat22}
{Patra}, S., {Evans}, Neal~J., I., {Kim}, K.-T., {Heyer}, M., {Kauffmann}, J., {Jose}, J., {Samal}, M.~R., \& {Das}, S.~R. 2022, {Tracers of Dense Gas in the Outer Galaxy}.
\newblock {\em \aj}, 164(4), 129.

\bibitem[{Pauly} and {Garrod}, 2018]{Pau18}
{Pauly}, T. \& {Garrod}, R.~T. 2018, {Modeling CO, CO$_{2}$, and H$_{2}$O Ice Abundances in the Envelopes of Young Stellar Objects in the Magellanic Clouds}.
\newblock {\em \apj}, 854, 13.

\bibitem[{Pety} et~al., 2005]{Pet05}
{Pety}, J., {Teyssier}, D., {Foss{\'e}}, D., {Gerin}, M., {Roueff}, E., {Abergel}, A., {Habart}, E., \& {Cernicharo}, J. 2005, {Are PAHs precursors of small hydrocarbons in photo-dissociation regions? The Horsehead case}.
\newblock {\em \aap}, 435(3), 885--899.

\bibitem[{Pietrzy{\'n}ski} et~al., 2013]{Pie13}
{Pietrzy{\'n}ski}, G., {Graczyk}, D., {Gieren}, W., {Thompson}, I.~B., {Pilecki}, B., {Udalski}, A., {Soszy{\'n}ski}, I., {Koz{\l}owski}, S., {Konorski}, P., {Suchomska}, K., {Bono}, G., {Moroni}, P.~G.~P., {Villanova}, S., {Nardetto}, N., {Bresolin}, F., {Kudritzki}, R.~P., {Storm}, J., {Gallenne}, A., {Smolec}, R., {Minniti}, D., {Kubiak}, M., {Szyma{\'n}ski}, M.~K., {Poleski}, R., {Wyrzykowski}, {\L}., {Ulaczyk}, K., {Pietrukowicz}, P., {G{\'o}rski}, M., \& {Karczmarek}, P. 2013, {An eclipsing-binary distance to the Large Magellanic Cloud accurate to two per cent}.
\newblock {\em \nat}, 495, 76--79.

\bibitem[{Rafelski} et~al., 2012]{Raf12}
{Rafelski}, M., {Wolfe}, A.~M., {Prochaska}, J.~X., {Neeleman}, M., \& {Mendez}, A.~J. 2012, {Metallicity Evolution of Damped Ly{$\alpha$} Systems Out to z \~{} 5}.
\newblock {\em \apj}, 755, 89.

\bibitem[{Rolleston} et~al., 2002]{Rol02}
{Rolleston}, W.~R.~J., {Trundle}, C., \& {Dufton}, P.~L. 2002, {The present-day chemical composition of the LMC}.
\newblock {\em \aap}, 396, 53--64.

\bibitem[{Rosolowsky} et~al., 2011]{Ros11}
{Rosolowsky}, E., {Pineda}, J.~E., \& {Gao}, Y. 2011, {Minimal HCN emission from molecular clouds in M33}.
\newblock {\em \mnras}, 415(2), 1977--1984.

\bibitem[{Ruffle} and {Herbst}, 2001]{Ruf01}
{Ruffle}, D.~P. \& {Herbst}, E. 2001, {New models of interstellar gas-grain chemistry - III. Solid CO$_{2}$}.
\newblock {\em \mnras}, 324, 1054--1062.

\bibitem[{Ruffle} et~al., 2007]{Ruf07}
{Ruffle}, P.~M.~E., {Millar}, T.~J., {Roberts}, H., {Lubowich}, D.~A., {Henkel}, C., {Pasachoff}, J.~M., \& {Brammer}, G. 2007, {Galactic Edge Clouds. I. Molecular Line Observations and Chemical Modeling of Edge Cloud 2}.
\newblock {\em \apj}, 671(2), 1766--1783.

\bibitem[{Russell} and {Dopita}, 1992]{Rus92}
{Russell}, S.~C. \& {Dopita}, M.~A. 1992, {Abundances of the heavy elements in the Magellanic Clouds. III - Interpretation of results}.
\newblock {\em \apj}, 384, 508--522.

\bibitem[{Schilke} et~al., 1997]{Schi97}
{Schilke}, P., {Groesbeck}, T.~D., {Blake}, G.~A., {Phillips}, \& {T.~G.} 1997, {A Line Survey of Orion KL from 325 to 360 GHz}.
\newblock {\em \apjs}, 108, 301--337.

\bibitem[{Seale} et~al., 2011]{Sea11}
{Seale}, J.~P., {Looney}, L.~W., {Chen}, C.-H.~R., {Chu}, Y.-H., \& {Gruendl}, R.~A. 2011, {The Evolution of Massive Young Stellar Objects in the Large Magellanic Cloud. II. Thermal Processing of Circumstellar Ices}.
\newblock {\em \apj}, 727, 36.

\bibitem[{Seale} et~al., 2009]{Sea09}
{Seale}, J.~P., {Looney}, L.~W., {Chu}, Y., {Gruendl}, R.~A., {Brandl}, B., {Chen}, C., {Brandner}, W., \& {Blake}, G.~A. 2009, {The Evolution Of Massive Young Stellar Objects in the Large Magellanic Cloud. I. Identification and Spectral Classification}.
\newblock {\em \apj}, 699, 150--167.

\bibitem[{Sewi{\l}o} et~al., 2022a]{Sew22a}
{Sewi{\l}o}, M., {Cordiner}, M., {Charnley}, S.~B., {Oliveira}, J.~M., {Garcia-Berrios}, E., {Schilke}, P., {Ward}, J.~L., {Wiseman}, J., {Indebetouw}, R., {Tokuda}, K., {van Loon}, J.~T., {S{\'a}nchez-Monge}, {\'A}., {Allen}, V., {Chen}, C. H.~R., {Hamedani Golshan}, R., {Karska}, A., {Kristensen}, L.~E., {Kurtz}, S.~E., {M{\"o}ller}, T., {Onishi}, T., \& {Zahorecz}, S. 2022,a {ALMA Observations of Molecular Complexity in the Large Magellanic Cloud: The N 105 Star-forming Region}.
\newblock {\em \apj}, 931a(2), 102.

\bibitem[{Sewi{\l}o} et~al., 2018]{Sew18}
{Sewi{\l}o}, M., {Indebetouw}, R., {Charnley}, S.~B., {Zahorecz}, S., {Oliveira}, J.~M., {van Loon}, J.~T., {Ward}, J.~L., {Chen}, C.-H.~R., {Wiseman}, J., {Fukui}, Y., {Kawamura}, A., {Meixner}, M., {Onishi}, T., \& {Schilke}, P. 2018, {The Detection of Hot Cores and Complex Organic Molecules in the Large Magellanic Cloud}.
\newblock {\em \apjl}, 853, L19.

\bibitem[{Sewi{\l}o} et~al., 2022b]{Sew22b}
{Sewi{\l}o}, M., {Karska}, A., {Kristensen}, L.~E., {Charnley}, S.~B., {Chen}, C. H.~R., {Oliveira}, J.~M., {Cordiner}, M., {Wiseman}, J., {S{\'a}nchez-Monge}, {\'A}., {van Loon}, J.~T., {Indebetouw}, R., {Schilke}, P., \& {Garcia-Berrios}, E. 2022,b {The Detection of Deuterated Water in the Large Magellanic Cloud with ALMA}.
\newblock {\em \apj}, 933b(1), 64.

\bibitem[{Shimonishi} et~al., 2016a]{ST16}
{Shimonishi}, T., {Dartois}, E., {Onaka}, T., \& {Boulanger}, F. 2016,a {VLT/ISAAC infrared spectroscopy of embedded high-mass YSOs in the Large Magellanic Cloud: Methanol and the 3.47 {$\mu$}m band}.
\newblock {\em \aap}, 585a, A107.

\bibitem[{Shimonishi} et~al., 2020]{ST20}
{Shimonishi}, T., {Das}, A., {Sakai}, N., {Tanaka}, K. E.~I., {Aikawa}, Y., {Onaka}, T., {Watanabe}, Y., \& {Nishimura}, Y. 2020, {Chemistry and Physics of a Low-metallicity Hot Core in the Large Magellanic Cloud}.
\newblock {\em \apj}, 891(2), 164.

\bibitem[{Shimonishi} et~al., 2021]{ST21}
{Shimonishi}, T., {Izumi}, N., {Furuya}, K., \& {Yasui}, C. 2021, {The Detection of a Hot Molecular Core in the Extreme Outer Galaxy}.
\newblock {\em \apj}, 922(2), 206.

\bibitem[{Shimonishi} et~al., 2008]{ST}
{Shimonishi}, T., {Onaka}, T., {Kato}, D., {Sakon}, I., {Ita}, Y., {Kawamura}, A., \& {Kaneda}, H. 2008, {AKARI Near-Infrared Spectroscopy: Detection of \ce{H2O} and \ce{CO2} Ices toward Young Stellar Objects in the Large Magellanic Cloud}.
\newblock {\em \apjl}, 686, L99--L102.

\bibitem[{Shimonishi} et~al., 2010]{ST10}
{Shimonishi}, T., {Onaka}, T., {Kato}, D., {Sakon}, I., {Ita}, Y., {Kawamura}, A., \& {Kaneda}, H. 2010, {Spectroscopic observations of ices around embedded young stellar objects in the Large Magellanic Cloud with AKARI}.
\newblock {\em \aap}, 514, A12.

\bibitem[{Shimonishi} et~al., 2012]{ST12}
{Shimonishi}, T., {Onaka}, T., {Kato}, D., {Sakon}, I., {Ita}, Y., {Kawamura}, A., \& {Kaneda}, H. 2012, {Akari Infrared Observations of Embedded YSOs in the Magellanic Clouds}.
\newblock {\em Publication of Korean Astronomical Society}, 27, 171--175.

\bibitem[{Shimonishi} et~al., 2013]{ST13}
{Shimonishi}, T., {Onaka}, T., {Kato}, D., {Sakon}, I., {Ita}, Y., {Kawamura}, A., \& {Kaneda}, H. 2013, {AKARI Infrared Camera Survey of the Large Magellanic Cloud. II. The Near-infrared Spectroscopic Catalog}.
\newblock {\em \aj}, 145, 32.

\bibitem[{Shimonishi} et~al., 2016b]{ST16b}
{Shimonishi}, T., {Onaka}, T., {Kawamura}, A., \& {Aikawa}, Y. 2016,b {The Detection of a Hot Molecular Core in the Large Magellanic Cloud with ALMA}.
\newblock {\em \apj}, 827b, 72.

\bibitem[{Shimonishi} et~al., 2023]{ST23}
{Shimonishi}, T., {Tanaka}, K. E.~I., {Zhang}, Y., \& {Furuya}, K. 2023, {The Detection of Hot Molecular Cores in the Small Magellanic Cloud}.
\newblock {\em \apjl}, 946(2), L41.

\bibitem[{Shimonishi} et~al., 2018]{ST18}
{Shimonishi}, T., {Watanabe}, Y., {Nishimura}, Y., {Aikawa}, Y., {Yamamoto}, S., {Onaka}, T., {Sakai}, N., \& {Kawamura}, A. 2018, {A Multiline Study of a High-mass Young Stellar Object in the Small Magellanic Cloud with ALMA: The Detection of Methanol Gas at 0.2 Solar Metallicity}.
\newblock {\em \apj}, 862, 102.

\bibitem[{Skillman} et~al., 1989]{Ski89}
{Skillman}, E.~D., {Kennicutt}, R.~C., \& {Hodge}, P.~W. 1989, {Oxygen Abundances in Nearby Dwarf Irregular Galaxies}.
\newblock {\em \apj}, 347, 875.

\bibitem[{Tielens} and {Hagen}, 1982]{Tie82}
{Tielens}, A.~G.~G.~M. \& {Hagen}, W. 1982, {Model calculations of the molecular composition of interstellar grain mantles}.
\newblock {\em \aap}, 114, 245--260.

\bibitem[{Turner} et~al., 1997]{Tur97}
{Turner}, B.~E., {Pirogov}, L., \& {Minh}, Y.~C. 1997, {The Physics and Chemistry of Small Translucent Molecular Clouds. VIII. HCN and HNC}.
\newblock {\em \apj}, 483(1), 235--261.

\bibitem[{Turner} et~al., 1999]{Tur99}
{Turner}, B.~E., {Terzieva}, R., \& {Herbst}, E. 1999, {The Physics and Chemistry of Small Translucent Molecular Clouds. XII. More Complex Species Explainable by Gas-Phase Processes}.
\newblock {\em \apj}, 518, 699--732.

\bibitem[{van Loon} et~al., 2008]{vanL08}
{van Loon}, J.~T., {Cohen}, M., {Oliveira}, J.~M., {Matsuura}, M., {McDonald}, I., {Sloan}, G.~C., {Wood}, P.~R., \& {Zijlstra}, A.~A. 2008, {Molecules and dust production in the Magellanic Clouds}.
\newblock {\em \aap}, 487, 1055--1073.

\bibitem[{van Loon} et~al., 2010a]{vanL10}
{van Loon}, J.~T., {Oliveira}, J.~M., {Gordon}, K.~D., {Meixner}, M., {Shiao}, B., {Boyer}, M.~L., {Kemper}, F., {Woods}, P.~M., {Tielens}, A.~G.~G.~M., {Marengo}, M., {Indebetouw}, R., {Sloan}, G.~C., \& {Chen}, C. 2010,a {A Spitzer Space Telescope Far-Infrared Spectral Atlas of Compact Sources in the Magellanic Clouds. I. The Large Magellanic Cloud}.
\newblock {\em \aj}, 139a, 68--95.

\bibitem[{van Loon} et~al., 2010b]{vanL10_b}
{van Loon}, J.~T., {Oliveira}, J.~M., {Gordon}, K.~D., {Sloan}, G.~C., \& {Engelbracht}, C.~W. 2010,b {A Spitzer Space Telescope Far-infrared Spectral Atlas of Compact Sources in the Magellanic Clouds. II. The Small Magellanic Cloud}.
\newblock {\em \aj}, 139b(4), 1553--1565.

\bibitem[{van Loon} et~al., 2005]{vanL05}
{van Loon}, J.~T., {Oliveira}, J.~M., {Wood}, P.~R., {Zijlstra}, A.~A., {Sloan}, G.~C., {Matsuura}, M., {Whitelock}, P.~A., {Groenewegen}, M.~A.~T., {Blommaert}, J.~A.~D.~L., {Cioni}, M., {Hony}, S., {Loup}, C., \& {Waters}, L.~B.~F.~M. 2005, {ESO-VLT and Spitzer spectroscopy of IRAS05328-6827: a massive young stellar object in the Large Magellanic Cloud}.
\newblock {\em \mnras}, 364, L71--L75.

\bibitem[{Wang} et~al., 2009]{Wan09}
{Wang}, M., {Chin}, Y.-N., {Henkel}, C., {Whiteoak}, J.~B., \& {Cunningham}, M. 2009, {Abundances and Isotope Ratios in the Magellanic Clouds: The Star-Forming Environment of N 113}.
\newblock {\em \apj}, 690, 580--597.

\bibitem[{Watanabe} and {Kouchi}, 2002]{Wat02}
{Watanabe}, N. \& {Kouchi}, A. 2002, {Efficient Formation of Formaldehyde and Methanol by the Addition of Hydrogen Atoms to CO in H$_{2}$O-CO Ice at 10 K}.
\newblock {\em \apjl}, 571, L173--L176.

\bibitem[{Watanabe} et~al., 2007]{Wat07}
{Watanabe}, N., {Mouri}, O., {Nagaoka}, A., {Chigai}, T., {Kouchi}, A., \& {Pirronello}, V. 2007, {Laboratory Simulation of Competition between Hydrogenation and Photolysis in the Chemical Evolution of H$_{2}$O-CO Ice Mixtures}.
\newblock {\em \apj}, 668, 1001--1011.

\bibitem[{Werner} et~al., 2004]{Wer04}
{Werner}, M.~W., {Roellig}, T.~L., {Low}, F.~J., {Rieke}, G.~H., {Rieke}, M., {Hoffmann}, W.~F., {Young}, E., {Houck}, J.~R., {Brandl}, B., {Fazio}, G.~G., {Hora}, J.~L., {Gehrz}, R.~D., {Helou}, G., {Soifer}, B.~T., {Stauffer}, J., {Keene}, J., {Eisenhardt}, P., {Gallagher}, D., {Gautier}, T.~N., {Irace}, W., {Lawrence}, C.~R., {Simmons}, L., {Van Cleve}, J.~E., {Jura}, M., {Wright}, E.~L., \& {Cruikshank}, D.~P. 2004, {The Spitzer Space Telescope Mission}.
\newblock {\em \apjs}, 154, 1--9.

\end{thebibliography}

\end{document}